\documentclass[prb,twocolumn]{revtex4-1} 

\usepackage{amsmath}  
\usepackage{amsfonts} 
\usepackage{graphicx} 

\begin{document}

\title{The Hobbyhorse of Magnetic Systems: The Ising Model}

\author{E. Ibarra-Garc\'ia-Padilla}
\email{eibarragp@fisica.unam.mx} 
\author{F. J. Poveda-Cuevas}
\affiliation{Instituto de F\'isica, Universidad Nacional Aut\'onoma de M\'exico,
Apartado Postal 20-364, M\'exico D.F. 01000, M\'exico.}

\author{C. G. Malanche-Flores}
\affiliation{\'Ecole polytechnique f\'ed\'erale de Lausanne (EPFL), CH-1015 Lausanne, Switzerland.}

\date{\today}

\begin{abstract}
 
The purpose of this article is to present a detailed numerical study of the second-order phase transition in the 2D Ising model. The importance of correctly presenting elementary theory of phase transitions, computational algorithms and finite-size scaling techniques results in a important understanding of both the Ising model and the second order phase transitions. In doing so, Markov Chain Monte Carlo simulations are performed for different lattice sizes with periodic boundary conditions. Energy, magnetization, specific heat, magnetic susceptibility and the correlation function are calculated and the critical exponents determined by finite-size scaling techniques. The importance of the correlation length as the relevant parameter in phase transitions is emphasized.
\end{abstract}

\maketitle 

\section{Introduction} 

When the Ising model was first introduced it appeared that the greatly over-simplified representation
of intermolecular forces on which this model is based would make it inapplicable to any real system. However, it seems that the essential features of cooperative phenomena do not depend on the details of intermolecular forces but on the mechanisms of propagation of long-range order, and the Ising model is a good first approximation to the problem \cite{Ising_history}. 

Nowadays it is the most-well studied model in statistical mechanics and although the Ising model is the simplest model used for describing a magnetic system it is also very versatile. For example, it can also be used to model the fluid critical point and binary alloy phase separation \cite{Tobochnik,Sethna} and variations of the Ising model have been used in high energy physics to explore the behavior of simple lattice gauge theories \cite{Tobochnik,Creutz}. 

Despite the simplicity of the model, it was solved analytically in the one dimensional case in 1925 by Ising \cite{Ising} and in 1944 for a two dimensional square lattice by Onsager \cite{Onsager}. It is worth to mention that while the 2D Ising model can be solved analytically, the 3D version does not have an analytical solution \cite{Istrail} or at least no solution has been found yet. However, solutions of the model can be found by numerical methods in $n$ dimensional square lattices without huge complications and for this reason it is usually presented as an exercise in textbooks \cite{Sethna, Thijssen, Krauth} and in some computational undergraduate courses.

Eventhough the computational algorithms for solving the Ising model are ``simple'' to program and are discussed and presented in textbooks \cite{Sethna, Thijssen, Krauth, Binder_book}, recovering thermodynamic properties and critical phenomena is not as straightforward as it might seem. It is our interest to give the reader a complete approach to the problem of critical phenomena in the 2D Ising model. In order to do so, elementary theory of phase transitions, computational algorithms and finite-size scaling techniques are presented.

This article is organized as follows. In section \ref{sec::magnetic} we introduce the Hamiltonian that is used to describe magnetic systems. From sections \ref{sec::phase} to \ref{sec::scaling_hyp} an introduction to phase transitions, the order parameter, the correlation function and the scaling hypothesis is made. Results from the mean-field and the exact solution are presented in section \ref{sec::mean_Onsager}, while in section \ref{sec::num_solution} the tools for the numerical solution are discussed. Later, in section \ref{sec::num_res} the numerical results are presented with their benchmarks and those results are discussed in section \ref{sec::discussion}, where the importance of the correlation length in phase transitions is emphasized. An example of the relevance of the correlation length is given in section \ref{sec::XY} where the XY model is discussed. Suggested problems are given in section \ref{sec::sugg_prob}.

\section{Magnetic systems}\label{sec::magnetic}

In describing the thermodynamics of magnetic systems a mean-field model called
the Weiss molecular field theory is used. The Hamiltonian of the system is the 
Heisenberg Hamiltonian given by \cite{Mattis_2,Heisenberg},
\begin{equation}\label{eq:Hamiltonian}
\mathcal{H} = - \frac{J}{2}\sum_{\langle ij \rangle} \textbf{s}_i \cdot \textbf{s}_j - \textbf{H} \cdot \sum_i \textbf{s}_i,
\end{equation}
where $\textbf{H}$ is the applied external magnetic field, $J$ is the coupling constant
which measures the strength of the interaction between spins or magnetic 
moments $\textbf{s}_i$. Ferromagnetic systems correspond to $J>0$ and
antiferromagnetic ones correspond to $J<0$. Each spin has a fixed location 
on a lattice and is labeled by the index $i$, whereas the index 
$\langle ij \rangle$ indicates that only nearest-neighbor interactions are 
considered. 

The net magnetization of the system is given by,
\begin{equation}\label{Magnetization}
\textbf{M} = \sum_i \textbf{s}_i.
\end{equation}

It is important to stress that setting $J$ as a constant in the Hamiltonian is a simplification.  The interaction between the spins $J_{ij}$ is named the exchange interaction and is an extremely complicated quantum phenomena, see for example Mattis's book \cite{Mattis_1}.

In the Ising model, the spin values are restricted to $\textbf{s}_i=\pm 1$. The
one dimensional Ising model does not exhibits a phase transition for $T>0$. 
On the other hand, the two dimensional Ising model solved 
by Onsager showed that in absence of an external magnetic field there is a 
continuous phase transition for finite temperature \cite{Onsager}.

\section{Phase transitions}\label{sec::phase}

The most familiar examples of phase transitions are those involving water, 
either melting ice to water or its evaporation. These phase transitions 
named first-order, are characterized by having discontinuous values
in its extensive variables\cite{footnote_extensive_var}. 
However, there is a subtlety when water is transformed into a gas. If it is at a 
sufficiently high temperature and pressure, there is no transition between 
a liquid and a gas. The limiting pressure and temperature above which there
is no phase transition are called the critical pressure and critical temperature,
respectively.

At the critical pressure and temperature there is a continuous phase
transition or second-order transition, and in this case the 
extensive variables of the system are continuous, while being its first derivatives,
as the specific heat or isothermal compressibility, discontinuous.
It is important to mention that this kind of phase transitions occur for all 
pure substances, not only water, but carbon dioxide, gold, and others. 
In a pressure-temperature phase diagram ($p-T$ diagram), this point is referred
as the critical point.  

Second order phase transitions are also seen in magnetic systems, such as the 
Curie point in ferromagnets, which separates the paramagnetic phase from 
the ferromagnetic one. That means that below the Curie temperature, the 
system presents spontaneous magnetization in absence of an external magnetic
field, whereas above it, the system is not magnetized and only responds when an 
external magnetic field is applied. It is our goal to illustrate what happens in second-order transitions and how do the thermodynamic properties of the systems behave.

\section{The order parameter and the correlation function}\label{sec::order_corr_function}

As stated above, in a second-order transition the first order
derivatives of the extensive variables of the system are discontinuous. These
thermodynamic quantities, such as the specific heat, the isothermal 
compressibility or the magnetic susceptibility in a magnetic system follow
a power law near the transition.

For example,
\begin{subequations}
  \begin{align}
    C_H \sim \left| 1- T/T_c \right|^{-\alpha},&  \\
    \chi_M \sim \left| 1- T/T_c \right|^{-\gamma},&
  \end{align}
\end{subequations}
where $C_H$ and $\chi_M$ denote the specific heat and the susceptibility, 
respectively. $T_c$ is the critical temperature, $T$ is the temperature 
at which the system is, $\alpha$ and $\gamma$ are critical exponents.

The dimension of space is very important in phase transitions. Critical exponents
fall into different universality classes depending upon both, the space 
dimension and on the system degrees of freedom. 
For example, the van der Waals equation of state 
reproduces correctly the critical exponents of a real liquid-gas for a four dimensional space.

Nonetheless, critical exponents give vital information about the thermodynamic 
system under study because they are linked with a key concept for 
understanding phase transitions, the order parameter. The relationship 
between the critical exponents and the order parameter is that the behavior of 
the last one near $T_c$ is usually described by the first ones. But what is the 
order parameter? In simple words the order parameter is the indicator of a 
phase transition. 

For example, in a magnetic system, the magnetization $M = \vert \textbf{M} \vert$
is the order parameter. Because in absence of an external magnetic field, in the
paramagnetic phase, the spins have equal probability of pointing in any 
direction, thus the net magnetic moment of the system vanishes $M=0$. 
Though, below  the critical temperature or Curie temperature, in the 
ferromagnetic phase, the spins tend to align in one direction, causing the 
net magnetic moment to be different from zero $M\neq0$. 
On the other hand, for the van der Waals fluid, the density is the order parameter
and for the Bose Einstein Condensation is the wavefunction itself, amongst
other examples. 

It is in the order parameter where the spontaneous symmetry breaking reflects. Considering the magnetic system, in the absence of an external magnetic field, the high-temperature phase (paramagnetic phase) exhibits an isotropic alignment of magnetic moments, but the low-temperature phase (ferromagnetic phase) does not. This isotropy is  broken ($M\neq0$), with a given preferential direction, called a spontaneous symmetry breaking.

To really understand the behavior of a system near a phase transition we 
need to look at its microscopic behavior. Such information is enclosed in the 
 correlation function $G(r,T)$, that expresses how the local order 
parameter at one position is correlated with itself a distance $r$ away. The correlation 
function is explicitly written as $G(r,T)$ to indicate that it depends both 
on the temperature of the system $T$ and the distance $r$.

Above the critical temperature, the correlation function falls off exponentially:
\begin{equation}\label{eq:cf_above}
G(r,T) \sim \exp\left(- \frac{r}{\xi(T)}\right) \qquad (T>T_c),
\end{equation} 
where $\xi(T)$ is called the correlation length, which as indicated, depends
on temperature. 

At the transition, the correlation falls off as a power law given by:
\begin{equation}\label{cf_critical}
G(r,T) \sim \frac{1}{r^{d-2 + \eta}} \qquad (T=T_c),
\end{equation}
where $d$ is the dimension of space and $\eta$ is a critical exponent. This 
power law decrease of the correlation function at the critical point implies that
there is no length scale in the system, and consequently far regions in the 
system are correlated.

Finally, below the critical temperature, the correlation function reaches a 
constant value for large $r$. Such ordering is called long-range order and it is
a consequence of cooperative effects that cause regions of space
to be correlated with nearby regions, which in turn causes a farther region to
be correlated. In this case, the deviation from the asymptotic value can be 
described by:
\begin{equation}\label{cf_below}
G(r,T)- G(\infty,T)\sim \exp\left(- \frac{r}{\xi(T)}\right) \qquad (T<T_c),
\end{equation}
The correlation length also follows a power law as the transition is approached
from either $T>T_c$ or $T<T_c$, given by:
\begin{equation}\label{cl_law}
\xi(T) \sim \left| 1- T/T_c \right|^{-\nu},
\end{equation}
being $\nu$  another critical exponent.

It is interesting to notice that the critical exponents are not independent from each other, because of the following scaling laws \cite{Huang}:
\begin{subequations}\label{eq::scaling_laws}
  \begin{align}
    \gamma &= \nu (2 - \eta),  \\
     2 &= \alpha + 2\beta + \gamma, \\
     \nu d &= 2 - \alpha, \\
     \gamma & = \beta (\delta -1),
  \end{align}
\end{subequations}
so it is only necessary to know two of them to determine the rest of them.

\section{The Scaling Hypothesis}\label{sec::scaling_hyp}

The scaling hypothesis is as its name indicates, a hypothesis. It does not rely on any model but has been very successful in correlating experimental data. The basic idea of the scaling hypothesis is that the long-range correlations around $T_c$ are responsible for all singular behavior \cite{Ma}.

So far, it seems that the important parameter in a phase transition is the order parameter and for a long time it was considered that if there was not a spontaneous symmetry breaking in the  order parameter in a system, then that system does not exhibit a phase transition. This belief is false and a brief example will be discussed later (see section \ref{sec::XY}).

In the scaling hypothesis, instead of looking at the order parameter, we focus our attention in a quantity we briefly mentioned in the last section: the correlation length $\xi$. It states that the divergence of $\xi$ near $T_c$ is responsible for the singular dependence on $1-T/T_c$ of physical quantities, and, as far as the singular dependence is concerned, $\xi$ is the only relevant length in the system \cite{Ma}.

It is not the scope of this paper to derive the scaling laws, neither to prove them by renormalization theory, but just to present the importance of the scaling hypothesis. An extensive discussion of the scaling hypothesis and renormalization theory can be found in references \cite{Ma,Wilson}. 

\section{Mean-Field and Onsager's Solution}\label{sec::mean_Onsager}

In the mean-field solution the Landau free energy \cite{Landau, Strecka, Baxter_book} is proposed as,
 \begin{equation}
F_L \approx  F_o + \frac{1}{2} a(T) N  m^2 + \frac{1}{4} b(T) N m^4 + \mathcal{O}(m^6),
\end{equation}
where $m=\vert \textbf{M} \vert/N$ and $N$ denotes the total number of lattices sites (spins). Under this proposal, if $a(T)$ and $b(T)$ are both positive, only $m=0$ is a minimum. On the other hand if $b(T)$ is positive and $a(T)$ changes sign, then $m=0$ is a local maximum and the minimum of $F_L$ occurs at $m \neq 0$, that is an indicator of a phase transition. The transition takes place at the critical temperature $T_c$, which is determined when $a(T)$ changes sign, i.e. $a(T_c) = 0$. As a result, $T_c$ for the Ising model is \cite{Baxter_book},
\begin{equation}
k_B T_c = z J,
\end{equation}
where $z$ is the number of nearest neighbors, so for the two dimensional case $k_B T_c = 4 J$. In the mean-field model it is also found that in absence of an external magnetic field,
\begin{subequations}
\begin{align}
&M = 0  &T \geq T_c,\\
&M \sim \left| 1- T/T_c \right|^{\beta}  &T<T_c,
\end{align}
\end{subequations}
with the critical exponent $\beta = 1/2$.  On the other hand, when $T=T_c$,
\begin{equation}
H \sim M^{\delta},
\end{equation}
with the critical exponent $\delta = 3$. 

On the other hand, although it is not the scope of this paper to present the exact solution, some important results that will be used in section \ref{sec::num_res} are given. The magnetization is found to be,
\begin{subequations}\label{eq::Onsager}
  \begin{align}
    &M = 0 & T > T_c,  \\
    &M = N \left[1 -  \sinh^{-4}(2 \beta J) \right]^{1/8} &  T<Tc,
  \end{align}
\end{subequations}
and the critical temperature is,
\begin{equation}
k_B T_c = \frac{2 J}{\ln(1 + \sqrt{2})} \approx 2.269 J.
\end{equation}
Also, near $T_c$ the heat capacity behaves as,
\begin{equation}\label{eq::CH_Onsager}
\begin{split}
C_H =&  N k_B \frac{2}{\pi} \left( \frac{2 J}{k_B T_c}\right)^2 \left[ -\ln \left( 1 - \frac{T}{T_c} \right)  \right.+ \\ & \left.  \ln \left( \frac{k_B T_c}{2J} \right) - \left( 1 + \frac{\pi}{4} \right) \right],
\end{split}
\end{equation}
so $C_H$ diverges logarithmically near the transition. From the results above, the critical exponents $\beta = 1/8$ and $\alpha = 0$ are found. All the critical exponents found in the mean-field and the exact solution are presented in TABLE \ref{Table::Exponents} in section \ref{sec::num_res}. A complete discussion on the exact solution can be found in references \cite{Huang,Strecka, Baxter_book}.

\section{Numerical Solution}\label{sec::num_solution}

\subsection{The Metropolis Hastings Algorithm}

In principle we could construct all the possible states the system can access $\{n\}$ and their energies $E_{\{n\}}$. With that information, we could construct the partition function and recover all the thermodynamics of the system. However there are $2^N$ possible states the system can access, so it is impractical to follow this path for large systems that obey $N \gg1$. This problem is solved by designing a Markov chain (or transition matrix) in such way that its stationary distribution is the desired distribution, in this case,
\begin{equation}\label{eq::prob_n}
P(\{n\}) = \frac{1}{Z} e^{-\beta E_{\{n\}}},
\end{equation}
where $\{n\}$ is a possible state of the system and $P(\{n\}) $ is its stationary distribution. 

But what is a Markov chain? It is one type of stochastic process. It is an evolution in time that is not determinist, but there is a transition probability from the current state to a new state. A Markov chain satisfies that the probability distribution of the next state depends only on the current state.

If an initial probability distribution is given $\textbf{P}(t=0)= \textbf{P}_0 $ and $\mathcal{P}$ is the transition matrix at one step, it is easy to prove that the probability distribution at the $n$-th step is simply,
\begin{equation}
\textbf{P}_{n} = \left(\mathcal{P}^N\right)^T \textbf{P}_{0},
\end{equation}
where the probability distributions are written as column vectors and the superindex $T$ denotes transposition.

It is our interest to know if there is a stationary distribution $\textbf{P}_{\infty}$, i.e. if the next limit exists,
\begin{equation}
\textbf{P}_{\infty} = \lim_{t\to\infty} \left( \mathcal{P}^T \textbf{P}_{0} \right).
\end{equation}
Because $ \textbf{P}_{0}$ is the initial distribution and does not depend on $t$, the problem reduces to determine if $ \lim_{t \to \infty} \mathcal{P}^T$ exists. It is straight forward to prove that if $ \textbf{P}_{\infty}$ exists and $ \textbf{P}_{0} =  \textbf{P}_{\infty}$, then $\forall \, t,  \textbf{P}_{t} =  \textbf{P}_{\infty}$.

A matrix is called ergodic if it is possible to go from every state to every other state (not necessarily in one move). This condition can be rewritten the following way:  A matrix is ergodic, i.e. $ \lim_{t \to \infty} \mathcal{P}^T$ exists if and only if its only unitary eigenvalue is 1. If the multiplicity of the eigenvalue is $m$, the theorem also holds. This condition is written as,
\begin{equation}
\textbf{P} =  \mathcal{P}^T \textbf{P},
\end{equation}
that expressed in index notation and using the fact that for a Markov chain $\sum_{i} \mathcal{P}_{ij}=1$, where $\mathcal{P}_{ij}$ is the transition probability from state $i$ to state $j$, is:
\begin{equation}
\sum_{j\neq i} P_i\,\mathcal{P}_{ij} = \sum_{j\neq i} P_j \, \mathcal{P}_{ji}.
\end{equation}
This equality is known as the global balance equations. However, if it is possible to found a stationary distribution that for all pair of states $i$ and $j$,
\begin{equation}\label{eq:detailed_balance}
 P_i\,\mathcal{P}_{ij} = P_j \, \mathcal{P}_{ji},
\end{equation}
then the global balance equations will trivially hold. This condition is named the detailed balance equation and is the basis of the Metropolis-Hastings algorithm.

In statistical mechanics, in the canonical ensemble, the stationary distribution is know and is given by the density matrix, so the probability that the system is in a state $\{n\}$ is given by eq. (\ref{eq::prob_n}). In that sense, in order to design a Markov chain that respects the desired distribution it is only necessary to consider the detailed balance equation (\ref{eq:detailed_balance}), that may be rewritten in a more enlightening manner as,
\begin{equation}
 \frac{P(\{n\})}{P(\{m\})} = \frac{\mathcal{P}(\{m\} \to \{n\})}{\mathcal{P}( \{n\} \to \{m\})}.
\end{equation}
And because of eq. (\ref{eq::prob_n}), this condition is simply,
\begin{equation}
 \frac{P(\{n\})}{P(\{m\})}  = \exp \left(-\beta \left[E(\{n\}) - E(\{m\}) \right] \right),
 \end{equation}
 so there is no need to calculate the partition function. 

The Metropolis-Hastings algorithm works as follows \cite{Metropolis,Hastings}:
\begin{enumerate}
\itemsep0em
\item The system is at state $\{m\}$ and the new configuration $\{ n \}$ is proposed with probability $g(\{m\} \to \{n\})$.
\item After this new state is proposed, it will accepted with probability $\alpha (\{m\} \to \{n\})$ or rejected with probability  $1 - \alpha (\{m\} \to \{n\})$. That way, the transition probability $\mathcal{P}(\{m\} \to \{n\})$  simply becomes,
\begin{equation}
\mathcal{P}(\{m\} \to \{n\}) = g(\{m\} \to \{n\}) \alpha(\{m\} \to \{n\}).
\end{equation}
\item The detailed balance equation is written as,
\begin{equation}
 \frac{P(\{n\})}{P(\{m\})} = \frac{g(\{m\} \to \{n\}) \alpha(\{m\} \to \{n\})}{g(\{n\} \to \{m\}) \alpha(\{n\} \to \{m\})},
\end{equation}
so, 
\begin{equation}
 \frac{\alpha(\{m\} \to \{n\})}{\alpha(\{n\} \to \{m\})} =  \frac{P(\{n\})}{P(\{m\})} \frac{g(\{n\} \to \{m\})}{g(\{m\} \to \{n\})},
\end{equation}
\item In order to fulfill the detailed balance equations, the acceptance distribution may be expressed as,
\begin{equation}
\alpha(\{m\} \to \{n\})= \min \left(1, \frac{P(\{n\})}{P(\{m\})} \frac{g(\{n\} \to \{m\})}{g(\{m\} \to \{n\})} \right).
\end{equation}
\end{enumerate}

In the Ising model when $\{n\}$ and $\{m\}$ only differ by one spin, $g(\{n\} \to \{m\}) = g(\{m\} \to \{n\})$, so the acceptance distribution is,
\begin{equation}
\alpha(\{m\} \to \{n\})= \min \left(1, \exp \left[-\beta \Delta E(\{n\}, \{m\}) \right] \right).
\end{equation}

With those ideas in mind, the Metropolis-Hastings algorithm for the Ising model is the following.
First, a $L \times L$ square lattice is created and in every site of the lattice
a spin is set with equal probability of being $\pm 1$. 

Each step of the algorithm is:
\begin{enumerate}
\itemsep0em 
\item Choosing randomly one site $k$ in the lattice.
\item Calculating the energy difference $\Delta E_k$ between
the actual energy and the energy if the spin is flipped.  
\item If $\Delta E_k < 0 $, we accept the new configuration. If not, we accept it
with probability $\exp\left(-\beta \Delta E_k \right)$ where $\beta= 1/k_B T$ and 
$k_B$ is the Boltzmann constant.
\item Energy and magnetization of the system are saved.
\end{enumerate}

The Metropolis-Hastings algorithm samples states according to the appropriate probability distribution and then temporal sequences of energies and magnetizations (generated by the sampling process) are then averaged and observables calculated. It is important to emphasize that the algorithm only generates configurations in agreement with the probability distribution, it does not compute thermodynamic quantities, but only samples ensembles from which thermodynamic quantities must be determined. In the next sections, techniques for recovering those quantities are discussed.

Before presenting those techniques, it is necessary to define a Monte Carlo step (MC step) as the product $N=L^2$ with the number of steps in the algorithm. For example, for a system with a lattice size $L=50$, a MC step is 2500 steps of the algorithm. That way a MC step takes into account the lattice size, while a step of the algorithm does not. In that sense, performing 1000 MC steps for different lattices sizes allows the same sampling in all the lattices, whereas performing 1000 steps of the algorithm for different lattices sizes does not. It is natural that the following question arises: How many MC steps produce a good sampling? This question is not so easy to answer because it depends on the temperature at which the system is as it will be clear later in section \ref{sec::num_res}. Nevertheless, if while computing observables, such as energy and magnetization, the curves are noisy, then it is necessary to perform more MC steps.

\subsection{Finite Size Scaling}

So far we have presented the mean-field, the Onsager and the numerical solution. However one problem arises in the numerical solution: while the mean-field and the exact solutions are in the thermodynamic limit ($N \to \infty$ and $V \to \infty$, but $N/V$ constant), in the numerical solution it is impossible to achieve $N,V \to \infty$.  Although we are incapable of achieving the thermodynamic limit a numerical computation, K. Binder developed the finite-size scaling technique for analyzing finite-size systems such as the ones considered in computational simulations~\cite{Finite_behavior,Binder_paper,Binder_Landau}.  

\begin{figure}[htbp]
	\centering
	\includegraphics[width=0.5\textwidth]{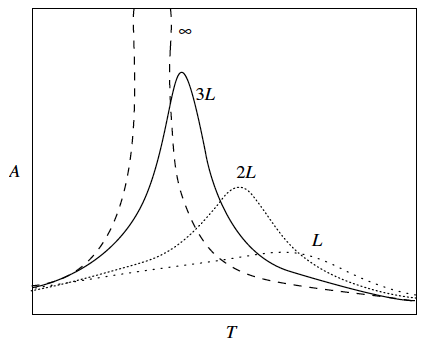}
	\caption{Typical behaviour of a physical quantity A vs temperature close to the critical point for various system sizes. Figure taken from Thijssen \cite{Thijssen}.}
	\label{fig::finite-size_scaling}
\end{figure}

As we have stated before, near the critical temperature, the correlation length diverges following a power law (\ref{cl_law}),
\begin{equation}
\xi \sim \left| 1- T/T_c \right|^{-\nu}.
\end{equation}
For a finite system, the thermodynamics quantities are smooth functions of the system parameters, so the divergences of the critical point phenomena are absent. Despite this fact, in the scaling region ($\xi >> L$), we can see traces of these divergences in the occurrence of peaks: peaks become higher and narrower and its location is shifted with respect to the location of the critical point as the system size increases (see FIG. \ref{fig::finite-size_scaling}). These characteristics of the peak shape as a function of temperature are described in terms of the so-called finite-size scaling exponents\cite{Thijssen}: 
\begin{itemize}
\itemsep0em
\item The shift in the position of the maximum with respect to the critical temperature is described by,
\begin{equation}
T_c(L) - T_c(\infty) \propto L^{-\lambda}
\end{equation}
\item The width of the peak scales as,
\begin{equation}
\Delta T(L) \propto L^{- \Theta}
\end{equation}
\item The peak height grows with the system size as,
\begin{equation}
A_{max}(L) \propto L^{\sigma_{max}}.
\end{equation}
\end{itemize}

Defining $t = \left| 1- T/T_c \right|$, the finite-size scaling Ansatz is formulated as follows \cite{Thijssen}:
\begin{equation}
\frac{A_L(t)}{A_\infty(t)} = f \left[ \frac{L}{\xi_{\infty}(t)}\right],
\end{equation}
where $A$ is a physical quantity. Assuming that the exponent of the critical divergence of $A$ is $\sigma$, and using the fact that  $\xi \sim t^{-\nu}$, the scaling Ansatz is formulated as,
\begin{equation}
A_L(t) = t^{-\sigma} f \left[ L t^{-\nu}\right],
\end{equation}
which can be rewritten as,
\begin{equation}\label{finite-size_scaling}
A_L(t) = L^{\sigma/ \nu} \phi \left[ L^{1/\nu} t\right],
\end{equation}
where the scaling function $f$ is replaced by $\phi$, by extracting the factor $(L t^\nu)^{\sigma/ \nu}$ from $f$ and writing the remaining function in terms of  $(L t^\nu)^{1/ \nu}$. From eq. (\ref{finite-size_scaling}) it is clear that \cite{Thijssen}:
\begin{itemize}
\itemsep0em
\item The peak height scales as $L^{\sigma/ \nu}$, hence $\sigma_{max}= \sigma/ \nu$.
\item The peak position scales as $L^{-1/ \nu}$, hence $\lambda = 1/ \nu$.
\item The peak width also scales as $L^{-1/ \nu}$, hence $\Theta = 1/ \nu$.
\end{itemize}
These are the finite-size scaling laws for any thermodynamic quantity which diverges at the critical point as a power law. From these laws it is clear that if the peak height, position and width are calculated as a function of the system size, the critical exponents $\nu$ and $\sigma$ can be determined.

Nevertheless, the finite-size scaling technique presents difficulties due a to phenomena named critical slowing-down \cite{Sethna,Thijssen,Krauth,Binder_book}. Because of the critical slowing-down, configurations change very slowly, and it is difficult to sample enough configurations. Near the critical point, the fluctuations increase and the time needed to obtain reliable values for the quantities measured also increases. As the system size increases, calculations for larger systems require more time, not only because of the computational effort needed per MC step for a larger system, but also because we need to generate more and more configurations in order to obtain reliable results.

\subsection{The correlation function}

In systems where a physical magnitude relies on position, one generally asks, given a measure at point $\textbf{r}_i$ what is the relation between another measure at a position $\textbf{r}_j$. This is given by the spatial correlation function and if the system presents translational and rotational symmetry (such as the Ising model), the correlation function does not depend on the absolute positions, but on the distance between them $r=\vert \textbf{r}_i - \textbf{r}_j \vert$. 
The correlation function we are interested in is the spin-spin correlation function that is given by, 
\begin{equation}\label{eq::corr_funct}
G(r,T) =  \langle s(0) s(r) \rangle - \langle s(0) \rangle ^2, 
\end{equation} 
where $\langle s(0) \rangle = \langle s(r) \rangle = M/N$ is the magnetization per site. Because of the fact that for a given temperature, $M$ reaches a constant value, the behavior of the correlation function is carried by the first term of eq. (\ref{eq::corr_funct}). Thus we will consider the correlation function only as,
\begin{equation}
G(r,T) =  \langle s(0) s(r) \rangle.
\end{equation}
We are limited to obtain the correlation function up to $L/2$, where $L$ is the lattice size. This came as a price of the periodic boundary conditions we are using. For example, if we were to calculate the correlation function up to the value $r=L$ we would find that the correlation function would be equal to 1 there, which is wrong because we would be computing the correlation function at $r=0$.

The process for numerically computing the correlation function is the following: For each spin in the lattice, we determine the value of the local correlation function in $r=n$ taking the average magnetic state of the nearest neighbors found advancing $n$ steps in one direction (not mixing $\hat{e}_i$ with $\hat{e}_j$, i.e. not moving in diagonals). The global correlation function is taken as the average of all the local correlation functions. The process is repeated for multiple simulations of the Ising model.

\subsection{Hints, tips and improvements to the algorithm}

As soon as Monte Carlo methods are used, one has to think on ways of making efficient calculations, as the brute force involved in a Monte Carlo simulation often requires a lot of trials to reduce standard deviation. 

\begin{itemize}
\itemsep0em
\item First of all, Monte Carlo methods are always good candidates for parallelization, which even in a dual-core cpu will half the calculation time. 
\item Second, one can remove some ``randomness'' to the method to improve efficiency. In the case of the Ising model, we know that the system will have preferred states as a function of the temperature. For temperatures below the critical temperature, once the system is near a local energy-minimum, it will hardly jump to another one (even if the energy gap is huge). All annealing methods are prone to this phenomena, and hence, once the standard deviations of the last steps start to decrease, the system should be randomized entirely to make sure we are not just sampling a single region of the entire space of states.
\item Third, one can also improve the selection of spins to flip, and change a completely uniform random distribution to a random walker (or walkers) that moves through the lattice. However, there are some considerations when doing this because we might actually modify the result or slow down the convergence rate. The step size is crucial to ensure a good convergence ratio. What could we say about the walker if the last ten attempts to switch a spin have failed? Well, the zone it is moving through is already in a stable position, hence, it sounds wise to increase the step size of the random walker. So the goal is to accelerate the relaxation process by trying to modify spins in unstable areas (which we assume are stable because a lot of changes are accepted). The best results are achieved if we monitor the standard deviation of the system in order to restart it as soon as it falls in a stable position.
\item Last but not least, it is important to perform simulations with a good random number generator. This is usually not a problem with most of computational languages, but it is something to take into account because a poor random generator will be unable to sample enough configurations, thus obtaining unreliable results. 
\end{itemize}

\section{Numerical results}\label{sec::num_res}

Energy, magnetization, specific heat and magnetic susceptibility are
studied first. It is a simple exercise to show that the specific heat $C_H$ and the 
magnetic susceptibility $\chi_M$ can be expressed in terms of the fluctuations of the
extensive variables,
\begin{subequations}
  \begin{align}
    C_H  &=  \left(\frac{\partial E}{\partial T}\right)_{H} = \frac{1}{k_B T^2} \left(\langle E^2 \rangle - \langle E \rangle^2 \right), \label{eq::CH}  \\
    \chi_M  &= \left(\frac{\partial M}{\partial H}\right)_{T} =  \frac{1}{k_B T} \left(\langle M^2 \rangle - \langle M \rangle^2 \right). \label{eq::chi_m}
  \end{align}
\end{subequations}

For a system in the thermodynamic limit, below the critical temperature, the system has equal probability of magnetizing with a positive value of $M$ or its negative value. Nevertheless, in the absence of an external magnetic field there is no chance for the system to go from the positive branch of spontaneous mangetization to the negative one and viceversa. However, for finite size systems there is a characteristic time (that depends of course on the lattice size) in which the system can go from one branch of the spontaneous magnetization to the other. Simply averaging the magnetization would yield an incorrect value, for this reason, instead of using $\langle M \rangle$ as the order parameter, $\langle \vert M \vert  \rangle$ will be used \cite{Binder_book}. If $\langle \vert M \vert \rangle$ is considered as the order parameter, $\chi_M$ is also calculated using $\langle \vert M \vert \rangle$ instead of $\langle  M \rangle$, so the susceptibility we compute is not exactly the ``true'' one.  However, in the $T<T_c$ region they are equal and in the $T>T_c$ region they only differ by a  constant, so the critical exponents are equal. Likewise, this consideration also ``smooths'' the computation of thermodynamic quantities that are derivatives of the order parameter, and in general of any physical quantity that presents a peak.

\begin{figure}[htbp]
	\centering
	\includegraphics[trim = 15mm 10mm 25mm 10mm, clip,width=0.5\textwidth]{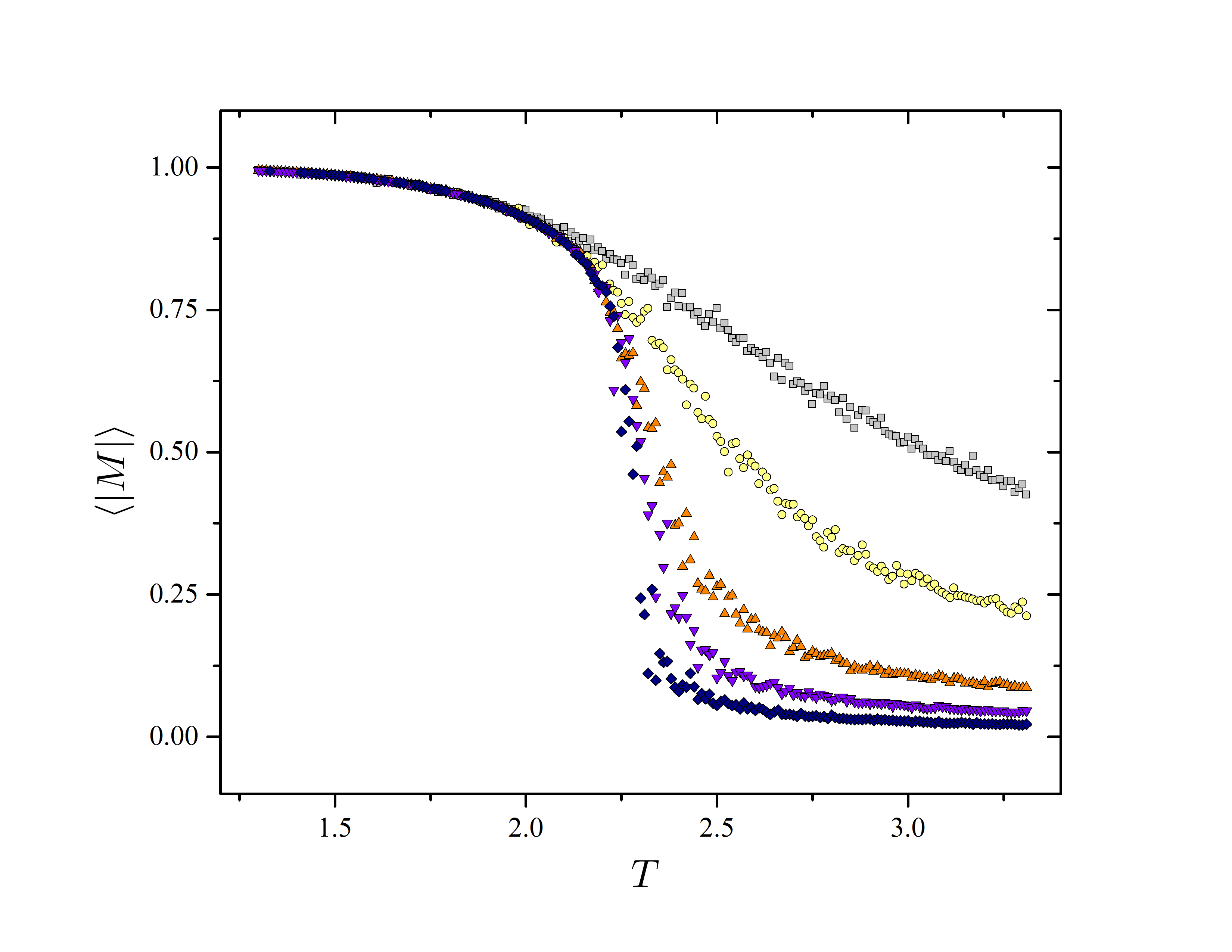}
	\caption{(Color online) Magnetization per site $M$ vs temperature $T$ for different lattice sizes $L$ = 5 (squares), 10 (circles), 25 (up triangles), 50 (down triangles) and 100 (diamonds) for the 2D Ising model.}
	\label{fig::effects_of_size_M}
\end{figure}

\begin{figure}[htbp]
	\centering
	\includegraphics[trim = 15mm 10mm 25mm 10mm,clip,width=0.5\textwidth]{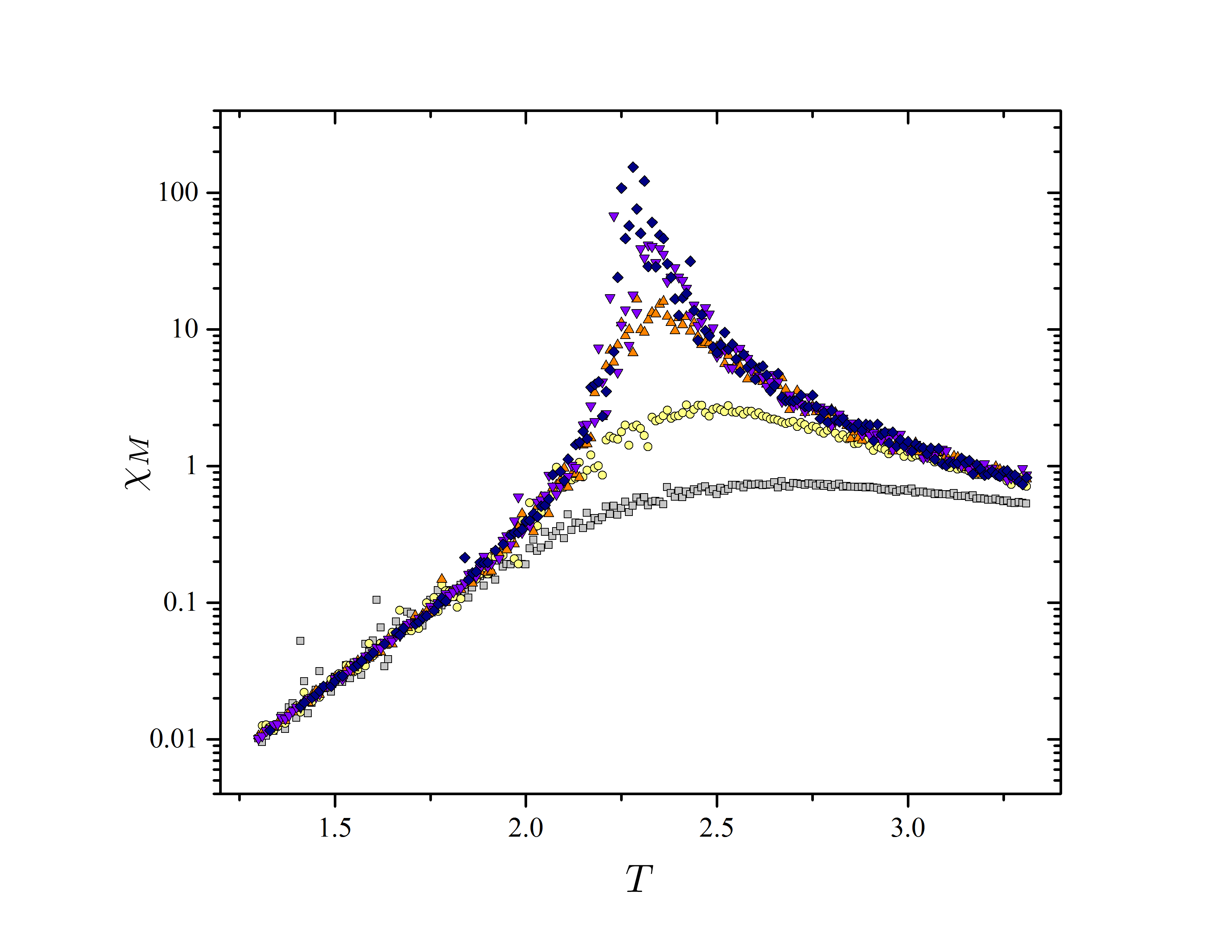}
	\caption{(Color online) Magnetic susceptibility per site $\chi_M$ vs temperature $T$ for different lattice sizes $L$ = 5 (squares), 10 (circles), 25 (up triangles), 50 (down triangles) and 100 (diamonds) for the 2D Ising model.}
	\label{fig::effects_of_size_Chi}
\end{figure}

\begin{figure}[htbp]
	\centering
	\includegraphics[trim = 15mm 10mm 25mm 10mm,clip,width=0.5\textwidth]{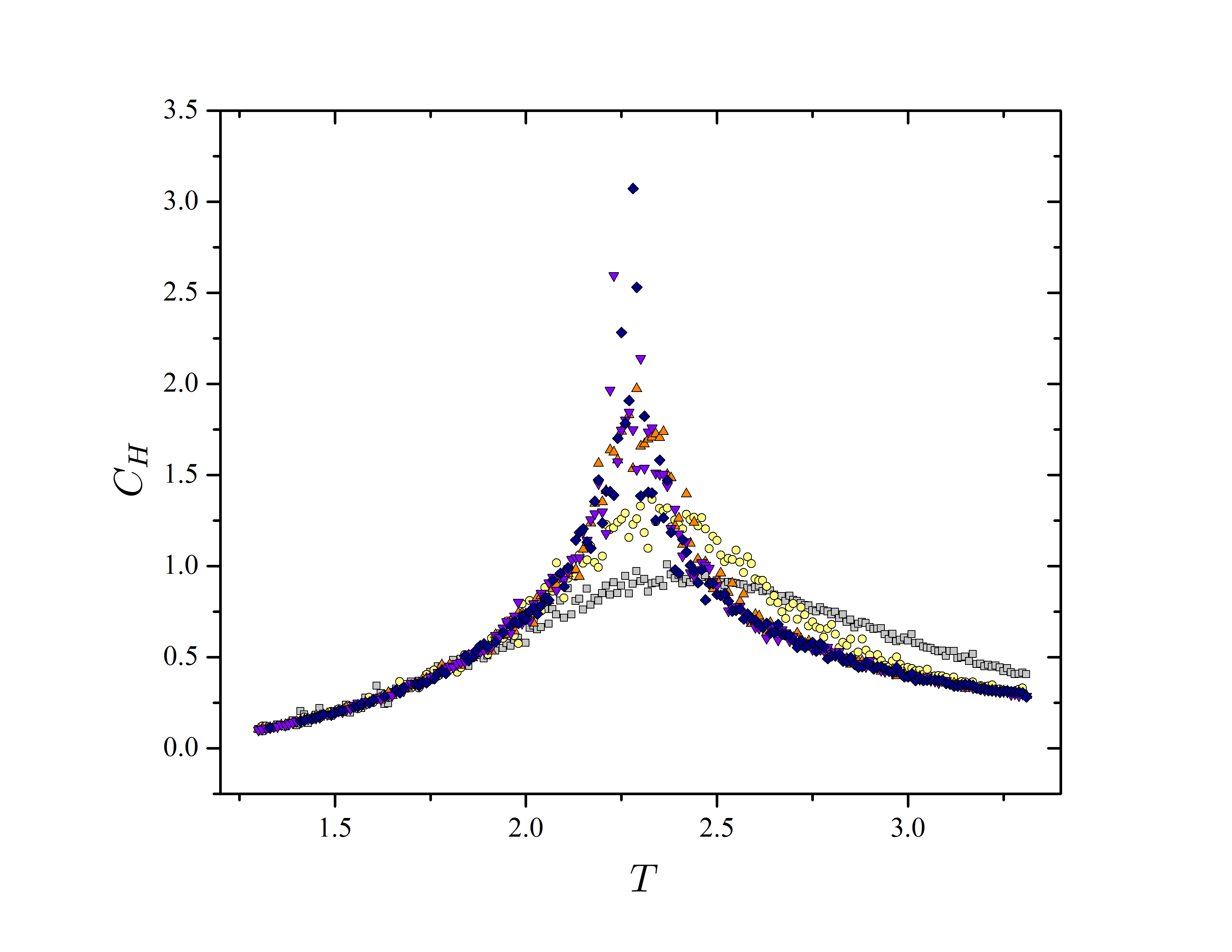}
	\caption{(Color online) Specific heat per site $C_H$ vs temperature $T$ for different lattice sizes $L$ = 5 (squares), 10 (circles), 25 (up triangles), 50 (down triangles) and 100 (diamonds) for the 2D Ising model.}
	\label{fig::effects_of_size_CH}
\end{figure}

Because magnetization $M$, energy $E$, specific heat $C_H$ and magnetic susceptibility $\chi_M$ are extensive variables, intensive ones were constructed by dividing them by the number of sites in the lattice $N=L^2$. From now on, $M$ will refer to magnetization per site, $E$ energy per site, $C_H$ specific heat per site and $\chi_M$ magnetic susceptibility per site. All simulations were performed considering $k_B=1$ and $J=1$. In FIGS. \ref{fig::effects_of_size_M},  \ref{fig::effects_of_size_Chi} and \ref{fig::effects_of_size_CH}, magnetization per site $M$, magnetic susceptibility per site $\chi_M$ and specific heat per site $C_H$ as functions of the temperature $T$ are presented. This plots were constructed by performing 15,000 MC steps per lattice size $L$. 

\subsection{How does the algorithm works}

\begin{figure}[htbp]
	\centering
	\includegraphics[trim = 15mm 10mm 20mm 10mm, clip,width=0.5\textwidth]{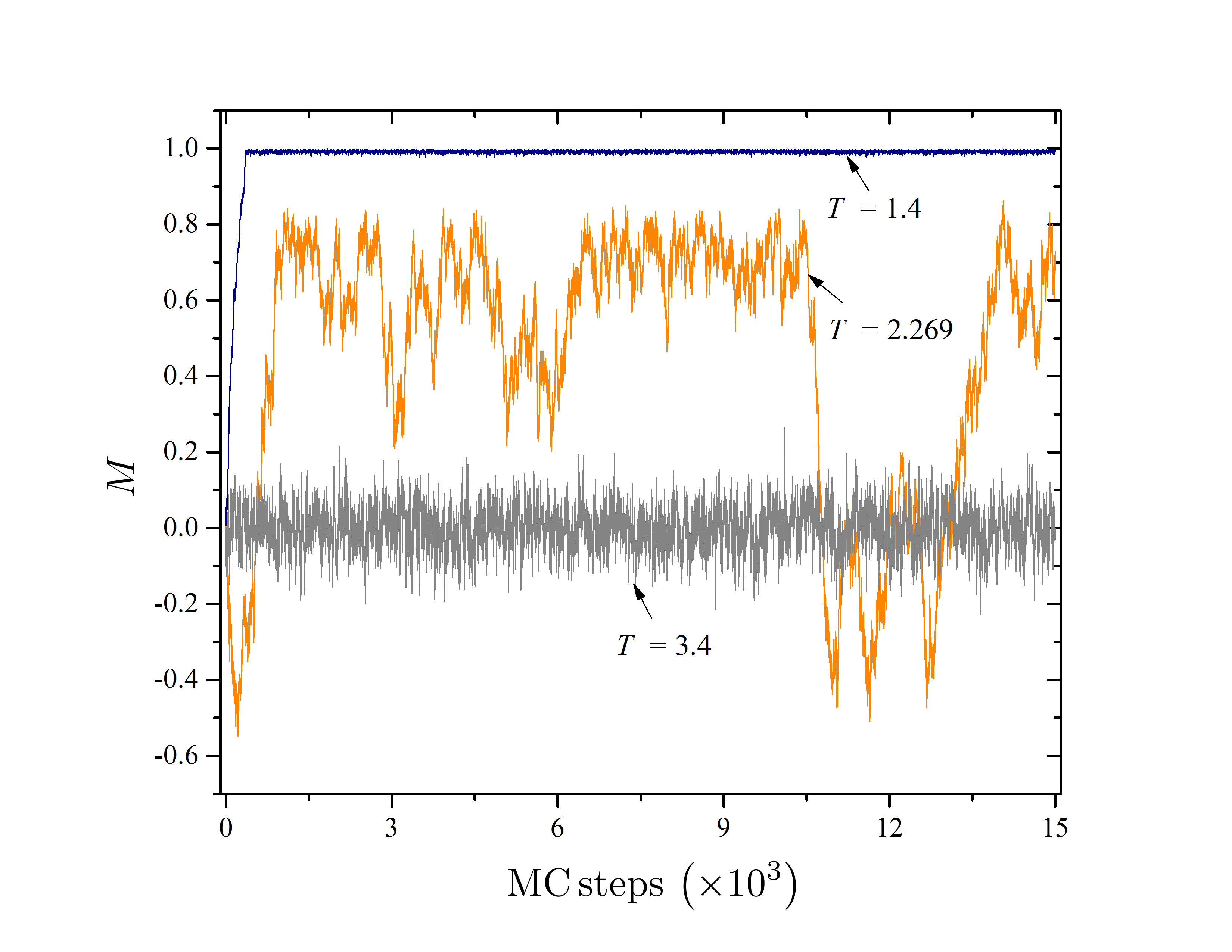}
	\caption{(Color online) Magnetization per site $M$ vs MC steps for temperatures $T$ = 1.4, 2.269 and 3.4 for a lattice size of $L=50$.}
	\label{fig::step}
\end{figure}

\begin{figure}[htbp]
	\centering
	\includegraphics[trim = 15mm 10mm 25mm 10mm, clip,width=0.5\textwidth]{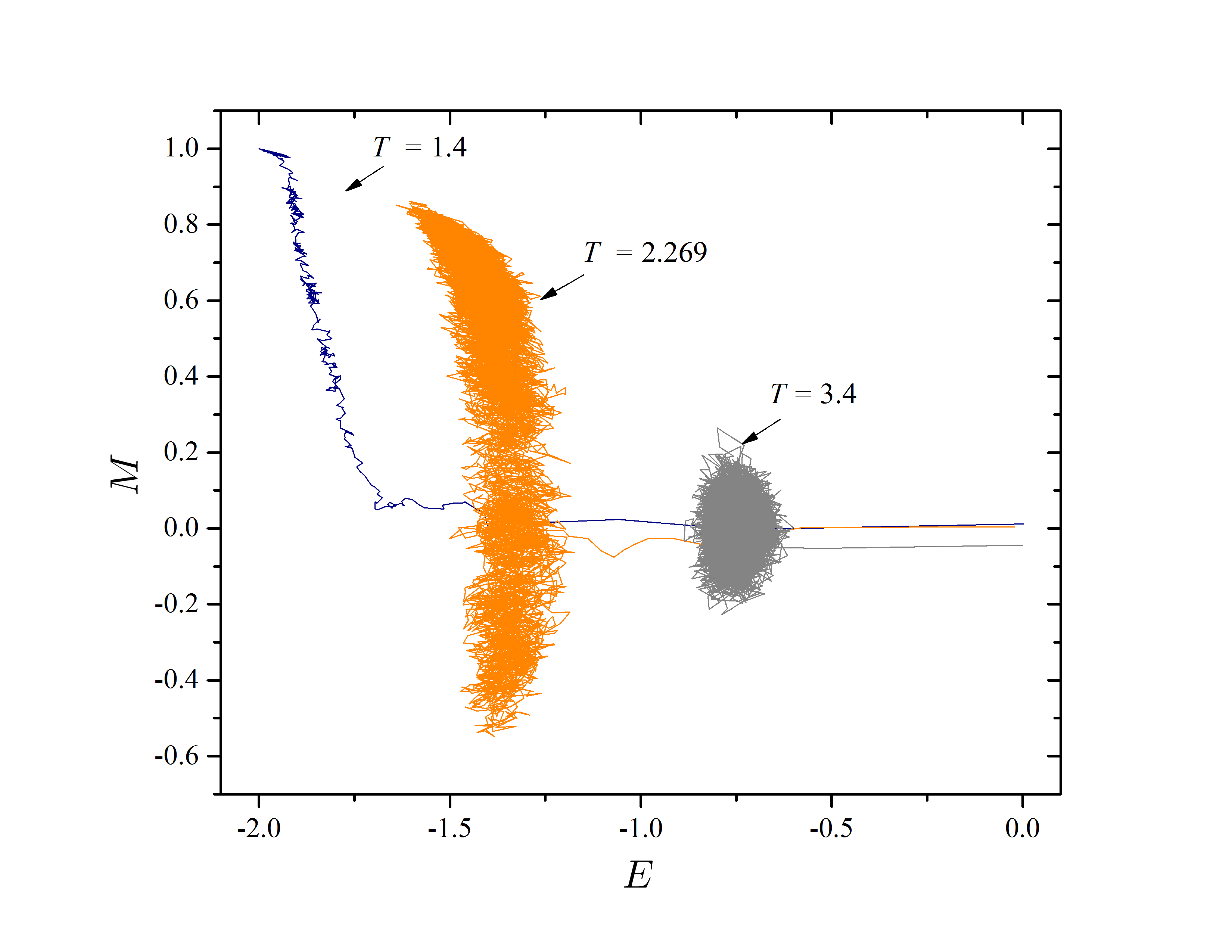}
	\caption{(Color online) Magnetization per site $M$ vs energy per site $E$ for temperatures $T$ = 1.4, 2.269 and 3.4 for a lattice size of $L=50$.}
	\label{fig::walk}
\end{figure}

In FIGS. \ref{fig::step} and \ref{fig::walk} simulations for a $L=50$ lattice are presented. In FIG. \ref{fig::step} the behavior of the order parameter is presented as a function of the 15,000 MC steps performed, while in FIG. \ref{fig::walk} it is presented as a function of the energy of the system.

From these figures important information is inferred. From FIG. \ref{fig::step} is clear that for $T<T_c$ and $T >T_c$ the behavior of the system is relatively simple. In the ferromagnetic phase the magnetization settles around a value $M \neq 0$, while in the paramagnetic phase the order parameter settles around $M=0$. But near the transition $T \approx T_c$, the system behaves exotically, the fluctuations in the order parameter are huge, $M$ does not settle around a concrete value, but covers a broad range of values. 

These observations are supported by FIG. \ref{fig::walk}, where it is clear that the algorithm samples states according to the probability distribution (see eq. (\ref{eq::prob_n})). 
\begin{itemize}
\itemsep0em
\item For $T<T_c$, there are few possible states to be sampled, but they are distributed in the phase space. This is due to the fact that states are sampled until the system magnetization reaches its final value. As the system gets magnetized, fewer and fewer states are accesible, because the acceptance probability gets smaller and smaller because $\Delta E_k < 0$ only if spins are flipped in the same direction of $M$ and $\exp\left(-\beta \Delta E_k \right)$ is very small, so flipping a spin in the opposite direction of $M$ is extremely improbable. 
\item For $T>T_c$ there are more possible states to be sampled, but they localized near $M=0$. This is obvious from the fact that the system is in its paramagnetic phase and fluctuations are only caused by thermal effects. 
\item Near $T_c$, the simulation samples a broad portion of the phase space. Near the scaling region generating statistically independent configurations consumes more computational time due to the critical slowing-down. This fact implies that near the transition configurations change very slowly, they correlate over large time scales.  For this reason, near the critical temperature more MC steps are needed than for $T<T_c$  and $T>T_c$.
\end{itemize}

\subsection{Critical exponents and the correlation function}

\begin{figure}[htbp]
	\centering
	\includegraphics[trim = 15mm 10mm 25mm 10mm, clip,width=0.5\textwidth]{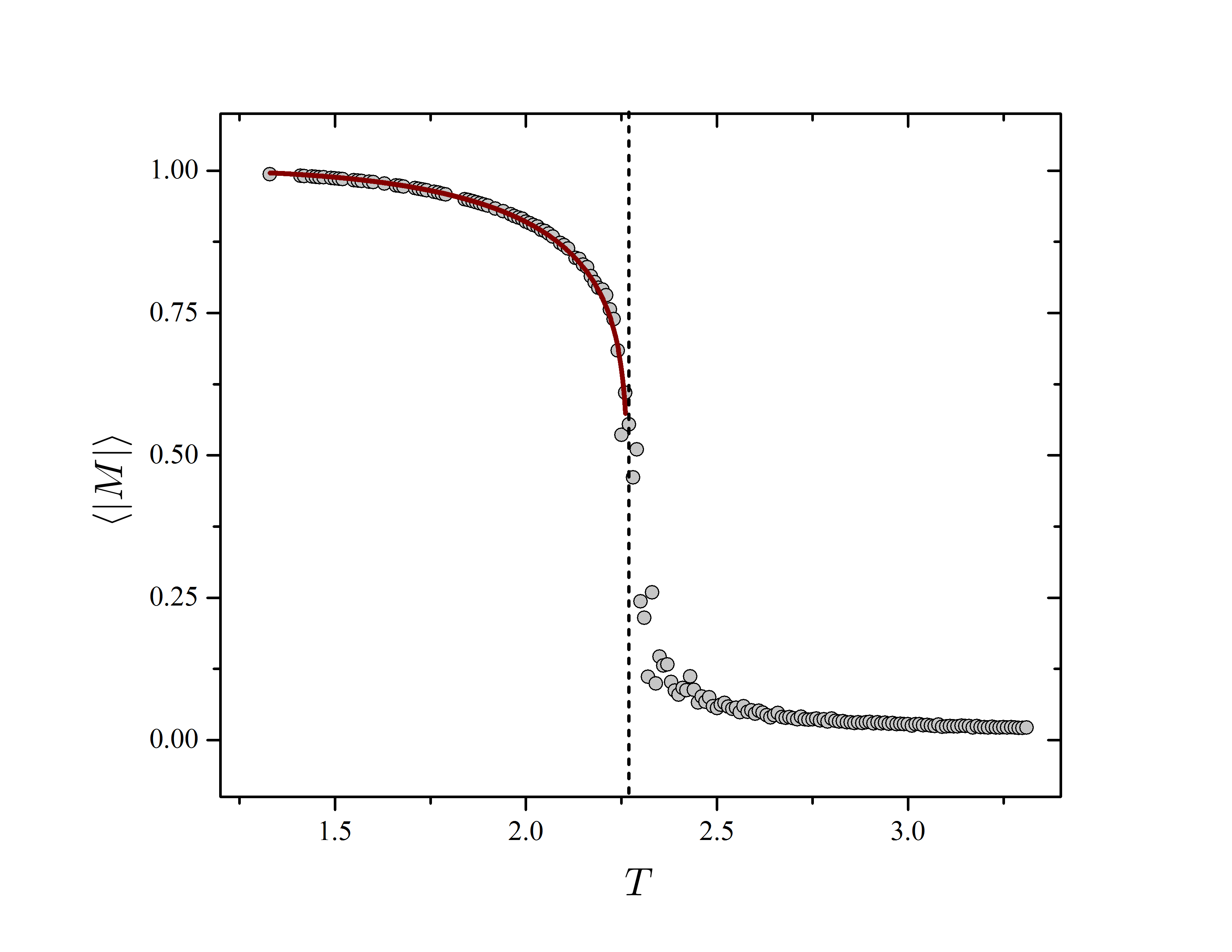}
	\caption{(Color online) Magnetization per site $M$ vs temperature $T$ for a lattice size of $L=100$ for the 2D Ising model. The curve in the left region was fitted with eq.(\ref{fit::Onsager}), where 
	$A = 1.020 \pm 0.026$, $B = 1.027 \pm 0.017$ and $C = 0.129 \pm 0.007$.}
	\label{fig::Magnetization_Ising}
\end{figure}

\begin{figure}[htbp]
	\centering
	\includegraphics[trim = 15mm 10mm 25mm 10mm,clip,width=0.5\textwidth]{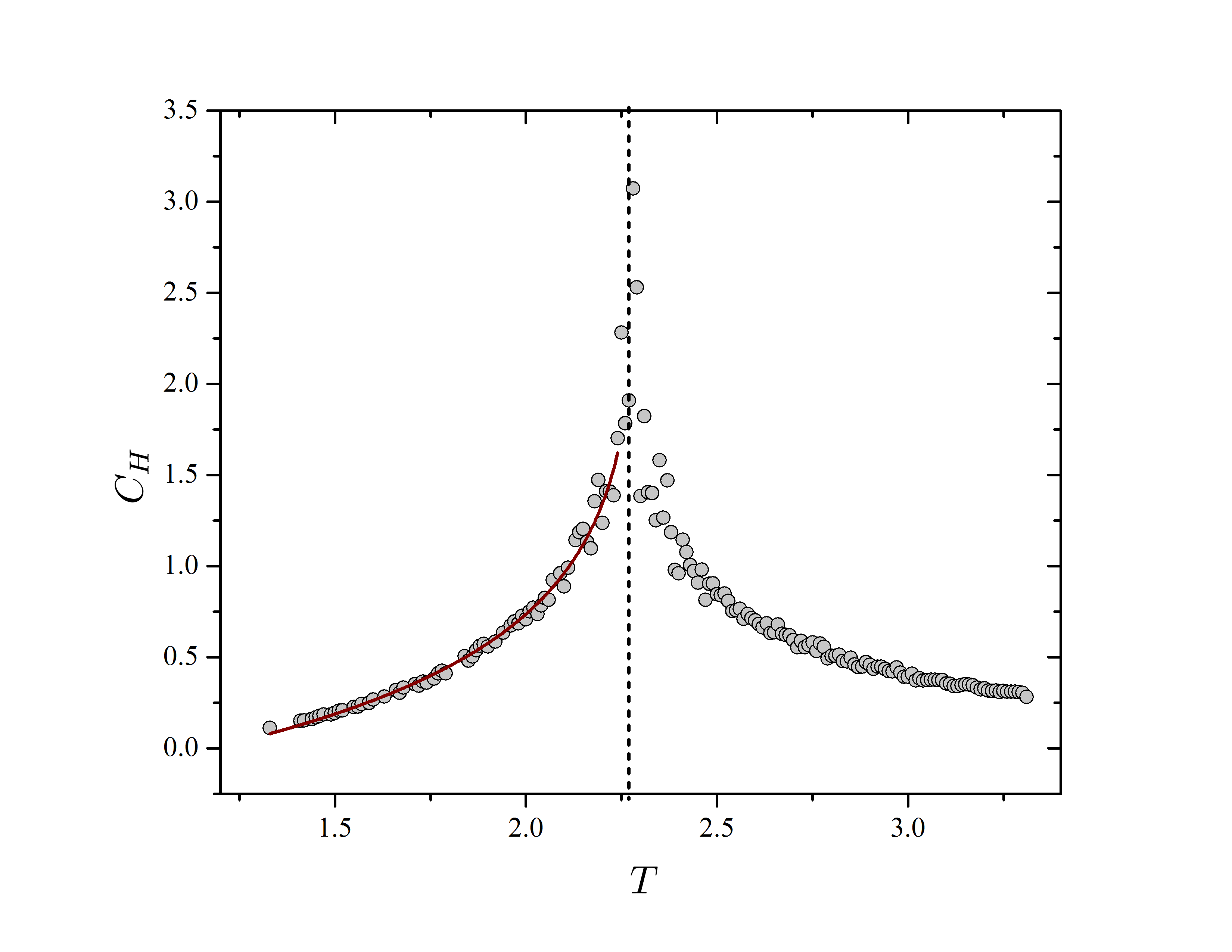}
	\caption{(Color online) Specific heat per site $C_H$ vs temperature $T$ for a lattice size of $L=100$ for the 2D Ising model. The curve in the left region was fitted with eq. (\ref{fit::log}), where $A = -0.404 \pm 0.030$, $B = 1.293 \pm 0.059$ and 
	$C = 2.303 \pm 0.011$.}
	\label{fig::CH_Ising}
\end{figure}

Despite the fact that for correctly determining the critical exponents, finite-size scaling techniques must be used and the correlation function studied, in the 2D Ising model there is an analytical solution. For this reason, we consider important to prove that our numerical solution is consistent with the analytical solution, hence FIGS. \ref{fig::Magnetization_Ising} and \ref{fig::CH_Ising} are fitted with Onsager's solution. The solution found by Onsager for the magnetization in the 2D Ising model is given by eq.~(\ref{eq::Onsager}), hence the curve used to fit the data presented in FIG. \ref{fig::Magnetization_Ising}  in the $T<T_c$ regime is,
\begin{equation}\label{fit::Onsager}
M = \left[A - B \sinh^{-4}\left(\frac{2}{T} \right) \right]^{C}.
\end{equation}
On the other hand, $C_H$ is given by eq. (\ref{eq::CH_Onsager}), so the curve used to fit the data presented in FIG. \ref{fig::CH_Ising} in the $T<T_c$ regime is,
\begin{equation}\label{fit::log}
C_H = A - B\ln \left( 1 - \frac{T}{C} \right).
\end{equation}
It is worth to notice the importance of considering large systems in order to obtain reliable results of the thermodynamic quantities because Onsager's solution is in the thermodynamic limit while numerical simulations are finite. In this manner, large systems were considered in order to ensure the correct behavior of magnetization and specific heat and from FIGS. \ref{fig::Magnetization_Ising} and \ref{fig::CH_Ising} the critical exponents $\beta$ and $\alpha$ are determined. 

\begin{figure}[htbp]
	\centering
	\includegraphics[trim = 15mm 10mm 25mm 10mm,clip,width=0.5\textwidth]{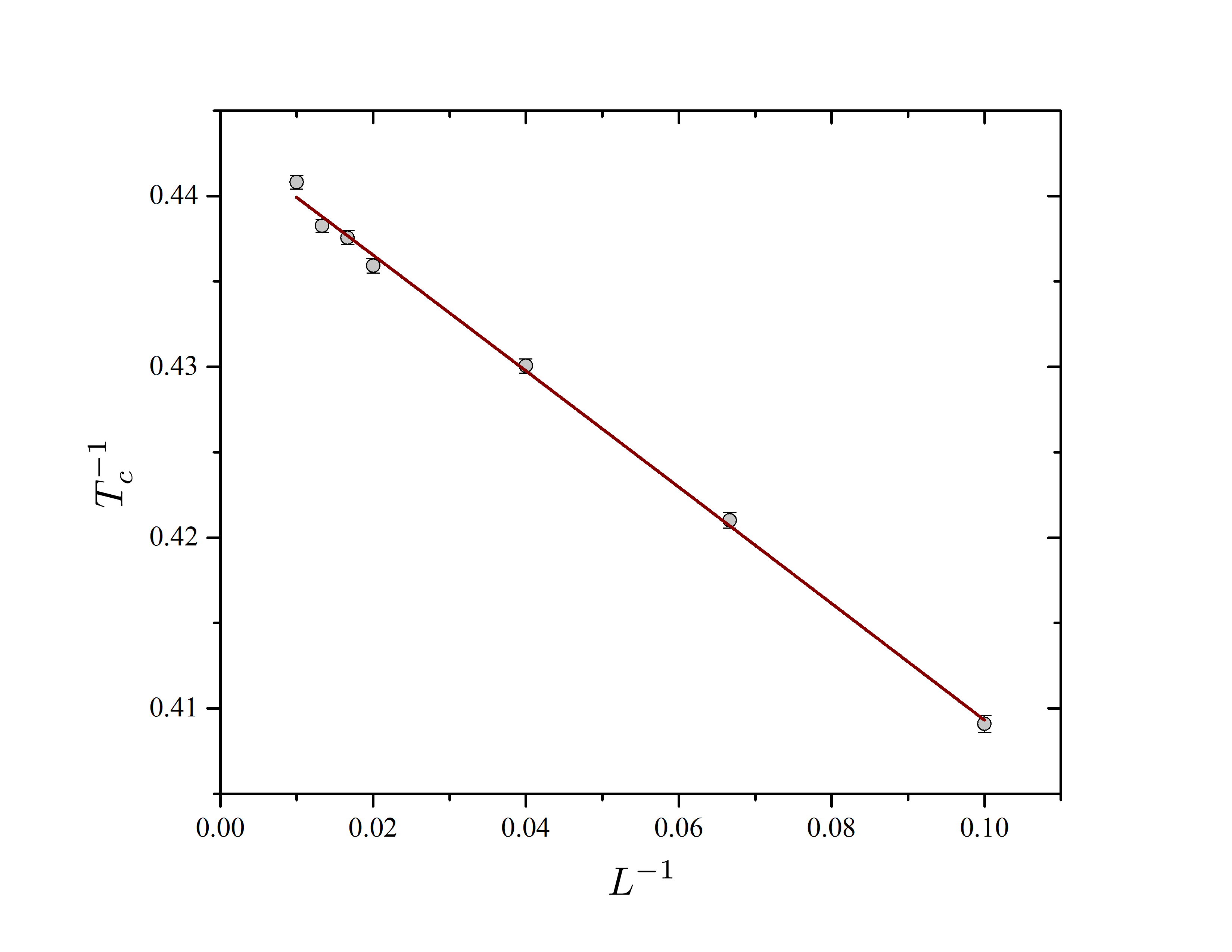}
	\caption{(Color online) Inverse of the critical temperature $T_c^{-1}$ vs inverse of lattice size $L^{-1}$ for the 2D Ising model. The curve was fitted with a power law $T_c^{-1} = T_{c_\infty}^{-1}- bL^{-1/\nu}$, and the critical exponent $\nu$ was determined.}
	\label{fig::nu_Ising}
\end{figure}

\begin{figure}[htbp]
	\centering
	\includegraphics[trim = 15mm 10mm 25mm 10mm,clip,width=0.5\textwidth]{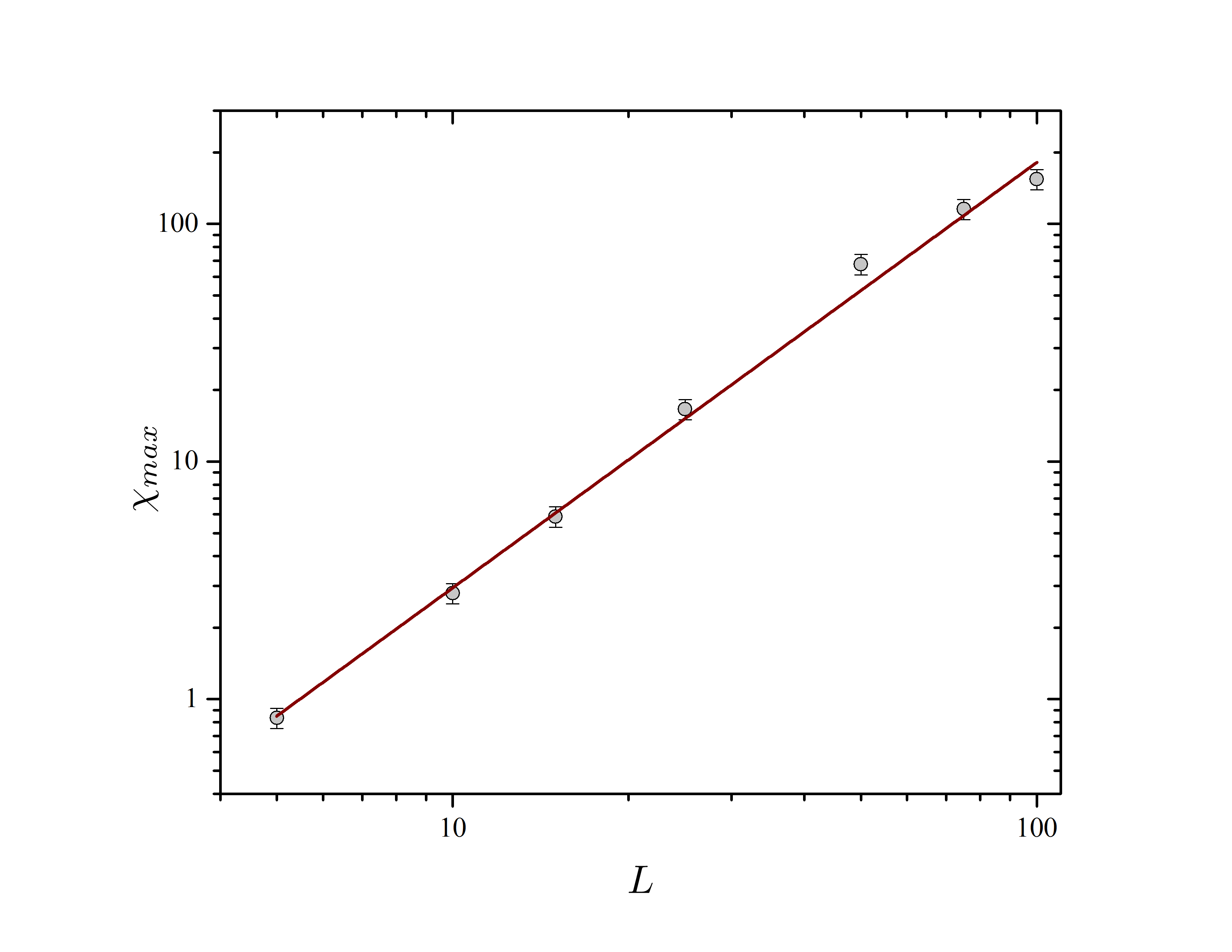}
	\caption{(Color online) Peak height of the magnetic susceptibility $\chi_{max}$ vs lattice size $L$ for the 2D Ising model. The curve was fitted with a power law $\chi_{max} = aL^{\gamma/\nu}$, and the critical exponent $\gamma$ was determined.}
	\label{fig::gamma_Ising}
\end{figure}

To determine $\nu$ and $\gamma$, FIGS. \ref{fig::nu_Ising} and \ref{fig::gamma_Ising} were constructed in order to use finite size-scaling techniques. In FIG. \ref{fig::nu_Ising} the critical temperature as a function of the lattice size $T_c(L)$ is obtained from FIG. \ref{fig::effects_of_size_M} by finding the inflection point in each curve. In order to construct FIG. \ref{fig::gamma_Ising}, 50 simulations of 15,000 MC steps each, were performed for every lattice size considered and the peak height of the magnetic susceptibility $\chi_{max}$ was registered for every repetition. These values were averaged and presented with its standard deviation of the mean. We know that near the critical temperature $T_c$ fluctuations are large, so the standard deviation of $\chi_{max}$ will be large too, but we are only interested in knowing how precise is its mean, that is why the standard deviation of the mean is used as a measure of uncertainty.

\begin{figure}[htbp]
	\centering
	\includegraphics[trim = 15mm 10mm 25mm 10mm, clip,width=0.5\textwidth]{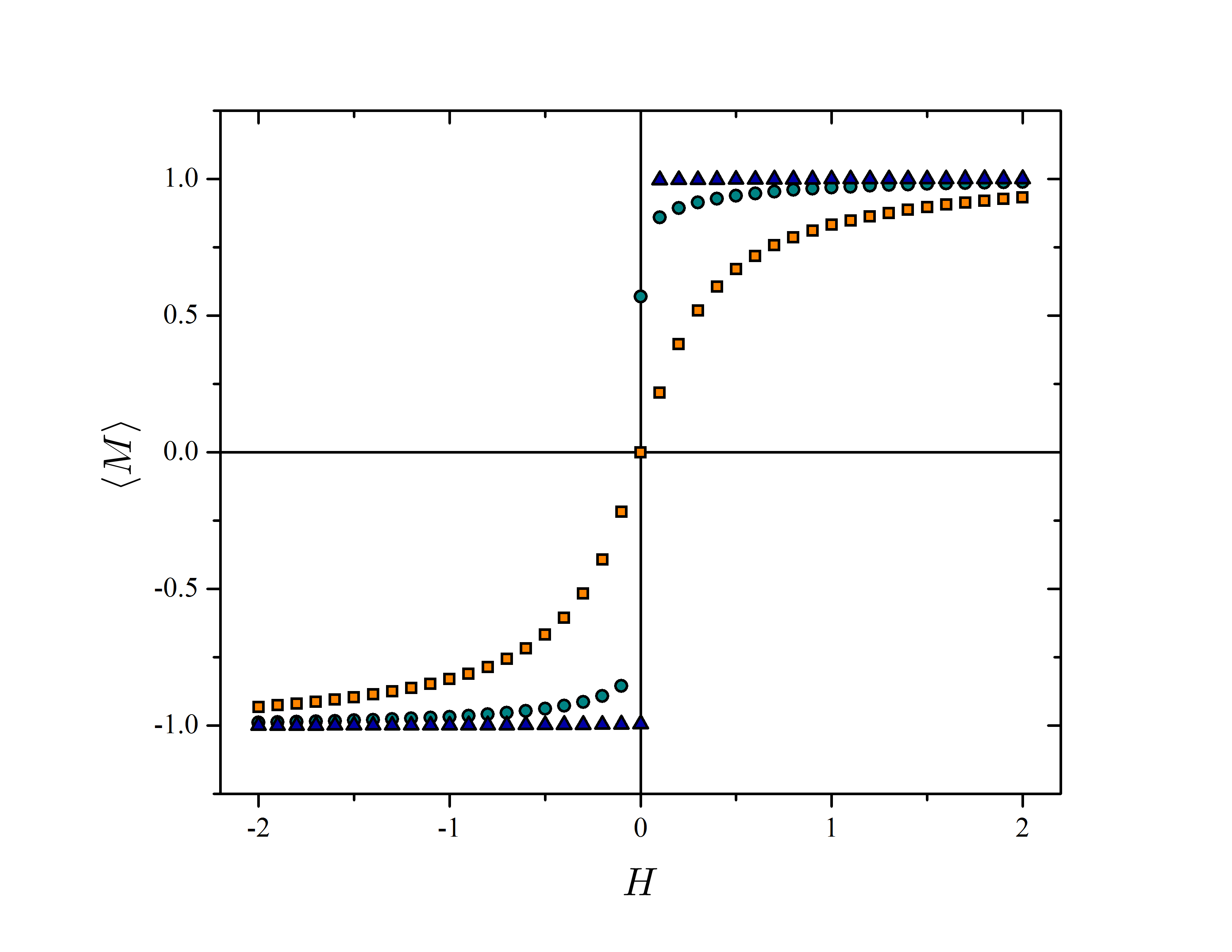}
	\caption{(Color online) Magnetization per site $M$ vs external magnetic field $H$ for temperatures $T$ = 1.3 (triangles), 2.269 (circles) and 3.3 (squares) for a lattice size of $L=50$.}
	\label{fig::MvsH}
\end{figure}

\begin{figure}[htbp]
	\centering
	\includegraphics[trim = 15mm 10mm 25mm 10mm, clip,width=0.5\textwidth]{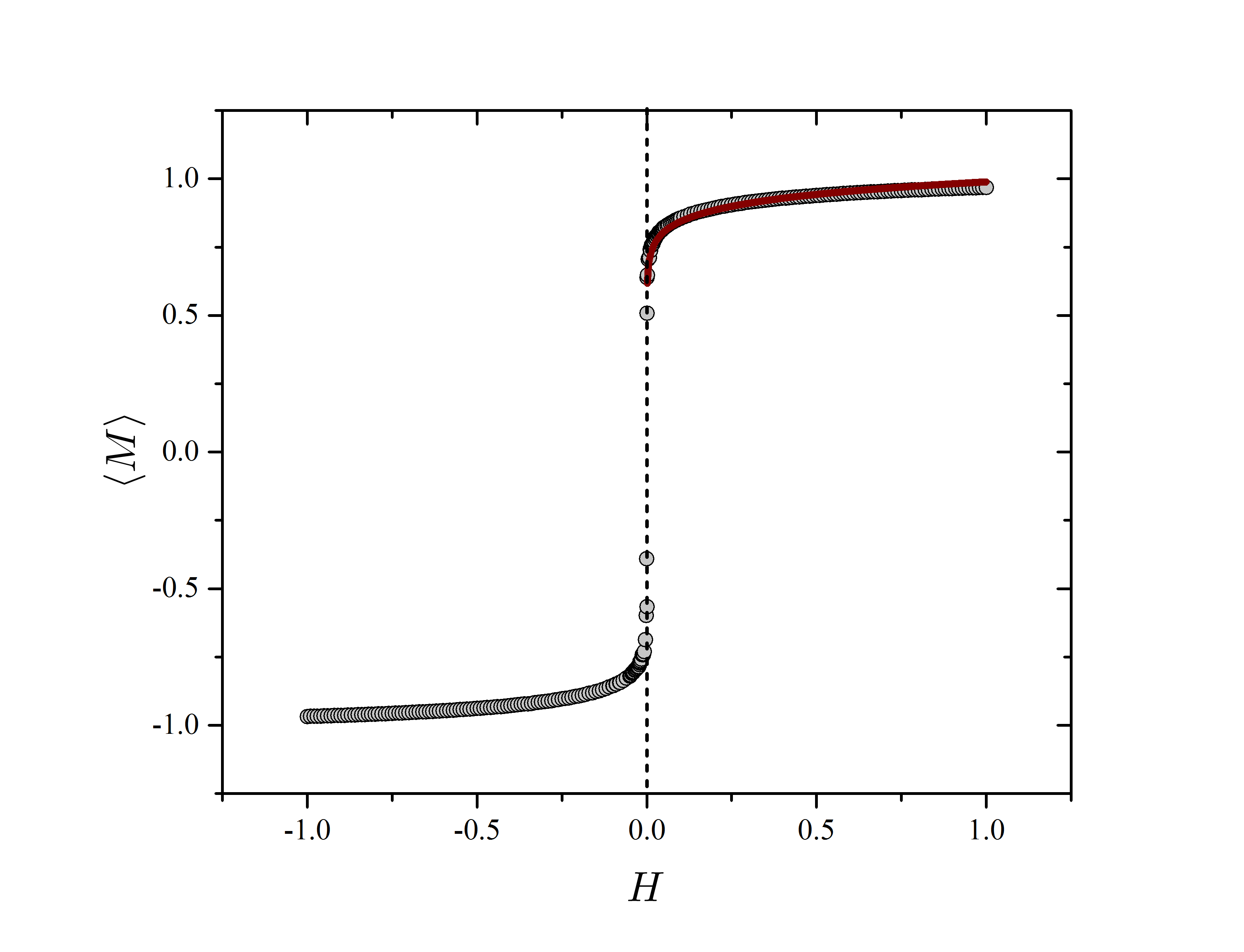}
	\caption{(Color online) Magnetization per site $M$ vs external magnetic field $H$ for a lattice size of $L=100$ at the critical temperature. The curve was fitted with a power law $M = a H^{1/\delta}$, and the critical exponent $\delta$ was determined.}
	\label{fig::M_delta_H}
\end{figure}

\begin{figure}[htbp]
	\centering
	\includegraphics[trim = 15mm 10mm 25mm 10mm, clip,width=0.5\textwidth]{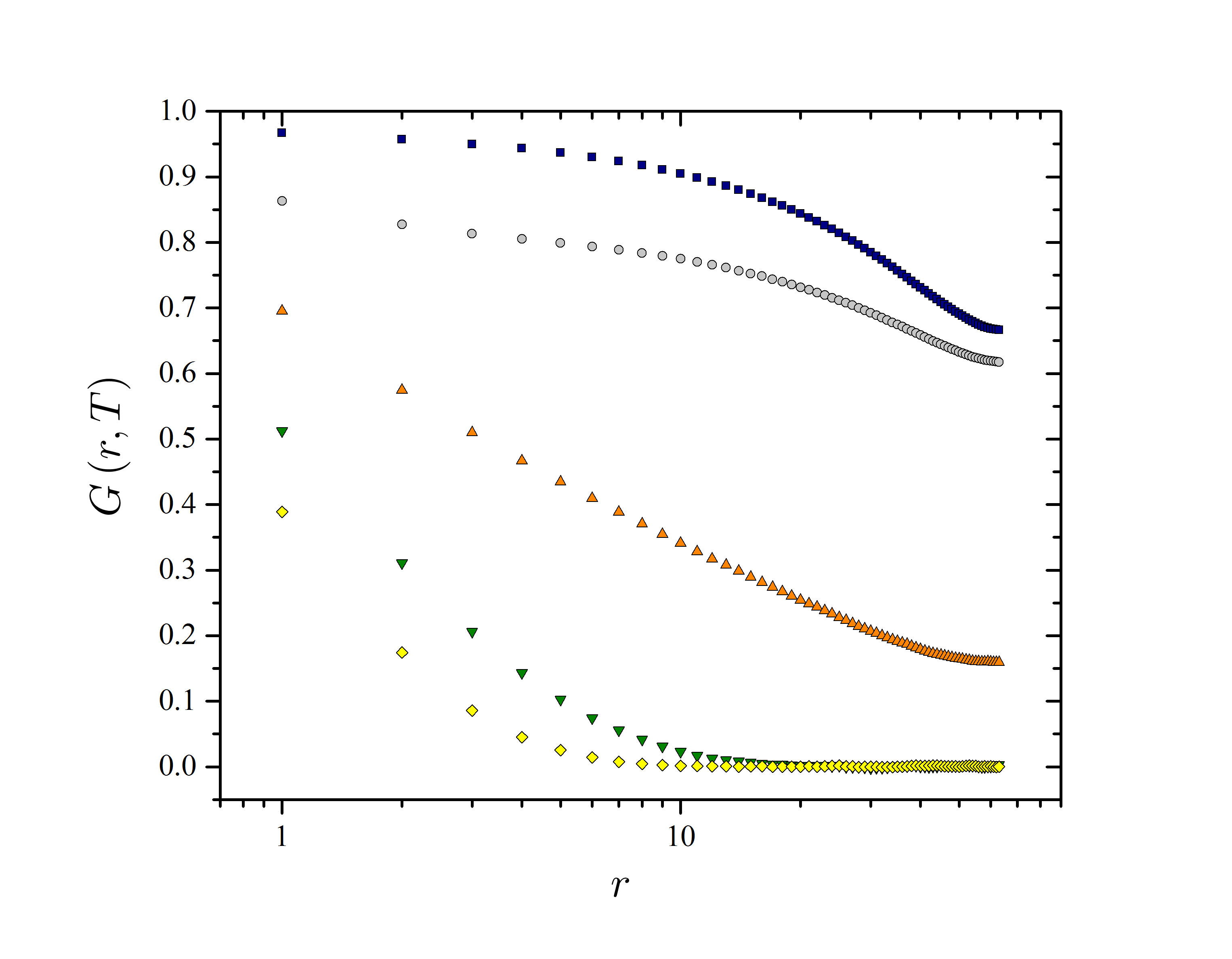}
	\caption{(Color online) Correlation function $G(r,T)$ for temperatures $T$ = 1.5 (squares), 2.0 (circles), 2.269 (up triangles), 2.6 (down triangles) and 3.1 (diamonds) for a lattice size of $L=128$.}
	\label{fig::correlation_T}
\end{figure}

\begin{figure}[htbp]
	\centering
	\includegraphics[trim = 15mm 10mm 25mm 10mm, clip,width=0.5\textwidth]{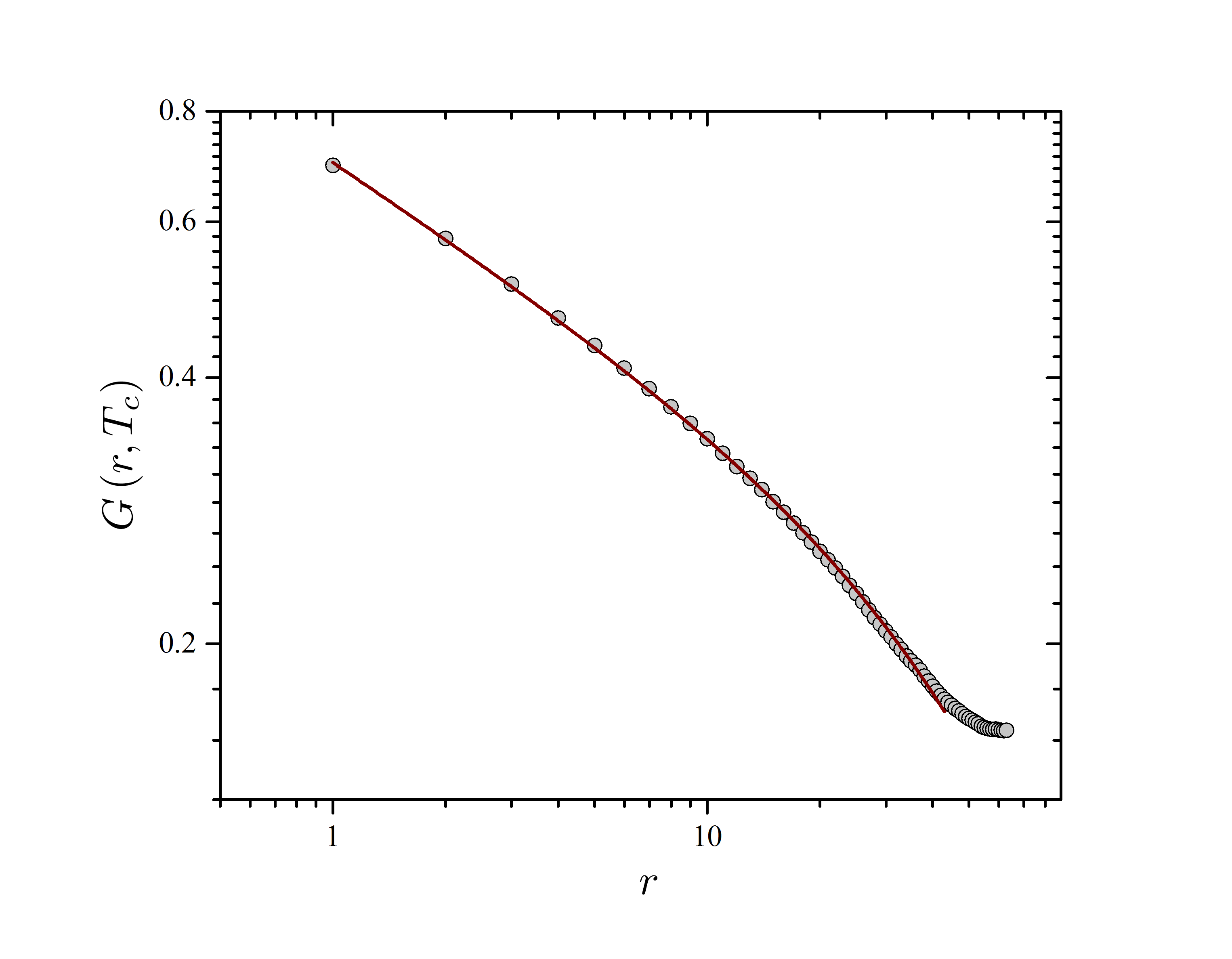}
	\caption{(Color online) Correlation function $G(r,T)$ at the critical temperature $T_c$ for a lattice size of $L=128$. The curve was fitted with eq.(\ref{fit::correlation}), where $A = 0.706 \pm 0.002$, $B = 108.525 \pm 2.752$ and $C = 0.277 \pm 0.002$. }
	\label{fig::correlation_eta}
\end{figure}

In FIG. \ref{fig::MvsH} we plot the magnetization $M$ as a function of the external magnetic field $H$ for different temperatures and from FIG. \ref{fig::M_delta_H} the critical exponent $\delta$ is determined. These plots were also constructed by performing 15,000 MC steps per lattice size $L$.
Later, in FIG. \ref{fig::correlation_T} the correlation function $G(r,T)$ is presented as a function of $r$ for different temperatures. The behavior discussed in section \ref{sec::order_corr_function} is clearly observed in this figure. For $T>T_c$ the correlation function falls of exponentially, at $T = T_c$ if follows a power law and for $T<T_c$, it reaches a constant value for large $r$.
Likewise, in FIG. \ref{fig::correlation_eta} the correlation function at the critical temperature $G(r,T_c)$ is presented as a function of $r$. The critical exponent $\eta$ is determined by fitting the data with the curve,
\begin{equation}\label{fit::correlation}
G(r,T_c) = A \frac{\exp\left( -r/B \right)}{r^C}.
\end{equation} 
These plots were constructed by performing 100 simulations of 15,000 MC steps each, for every temperature considered and the correlation function was registered for every repetition and then averaged.

Finally, in TABLE \ref{Table::Exponents}, the critical exponents for the two dimensional Ising model are presented as found by Onsager and the mean-field and numerical solutions.

\begin{table}[htbp]
\centering
\caption{Critical exponents and critical temperature of the 2D Ising model given by Onsager, the mean-field and the numerical solutions.}
\begin{ruledtabular}
\begin{tabular}{c | c c c }
\hline
Exponent & Onsager & Mean-Field & Numerical \\
\hline
$\alpha$ & $0$ & 0 & 0   \\
$\beta$ & $0.125$ & 0.5 & 0.129 $\pm$ 0.007 \\
$\gamma$ & $1.750$ & 1 & 1.779 $\pm$ 0.225 \\
$\nu$ & $1$ & 0.5 & 0.994 $\pm$ 0.098   \\
$\eta$ & $0.250$ & 0 & 0.277 $\pm$ 0.002  \\
$\delta$ & $15$ & 3 & 14.641 $\pm$ 0.821 \\
\hline
$T_c$ & 2.269 & 4 & $2.269 \pm 0.002$ \\
\end{tabular}
\end{ruledtabular}
\label{Table::Exponents}
\end{table}

\section{Discussion}\label{sec::discussion}

From TABLE \ref{Table::Exponents} is clear that the mean-field solution is inconsistent with the Onsager solution. Why does the mean-field solution is not consistent? In its derivation, fluctuations in the order parameter are neglected. However, in the scaling region, these fluctuations are of great relevance to the thermodynamic quantities, so neglecting them will of course generate incorrect results in this region, such as the critical temperature and the critical exponents. 

The main problem with most of the mean-field solutions is that the Landau free energy coefficients are expanded in Taylor series around $T_c$. This expansion presupposes that the coefficients are analytic around $T_c$, a supposition that is not valid. It is also worth to notice that the mean-field model presented here gives the same critical exponents for any dimension. This of course is a severe problem, because universality classes depend both on the space dimension and on the system degrees of freedom. Although most of the mean-field models do not reproduce correctly the critical exponents, they provide a useful insight to phase transitions, so they should not be discarded. Actually mean-field models such as the Ginzburg-Landau model, are used for describing phenomena such as superfluidity and superconductivity, see for example Annett's book \cite{Annett}. 
On the other hand, the numerical results are consistent with the analytic solution and the critical exponents determined by the numerical simulation obey the scaling laws (eqs. (\ref{eq::scaling_laws})). These are good indicators that numerical solutions are reliable for recovering information about the thermodynamics of a system.

In the numerical solutions, if very small systems are considered, then no reliable information can be inferred (see for example the $L= 5$ curve in FIG. \ref{fig::effects_of_size_M}). And although the thermodynamic limit is impossible to achieve in numerical simulations, finite-size scaling techniques provide a useful tool to recover information about the critical exponents. This technique requieres to consider large systems in order to obtain solid information, but as we have stated before, in the scaling region, the critical slowing-down phenomena is present, so sampling enough configurations requieres a lot of MC steps. This problem is partially solved by using other algorithms that respect the detailed balance equations, but that improve the effectiveness of the algorithm update (instead of flipping one spin, clusters are flipped) so they reduce the critical slowing-down phenomena~\cite{Wolff_PRL,Wolff_DESY,Barkema}.

So far we have seen that two particular situations emerge in the scaling region: the mean-field solution fails to reproduce the critical exponents and the numerical solution exhibits critical slowing-down. These situations illustrate the fact that in experiments is very hard to set the system at the critical point, it ``does not like to be'' at the critical point, so critical phenomena (such as critical opalescence) is hard to achieve. This difficulty to be at the critical point is exhibited in the mean-field solution in the fact that performing an analytic expansion around the critical point fails; and in the numerical solution in the critical slowing-down, where is difficult to sample enough configurations: the system does not ``feel comfortable'' in the critical region. This ``unconformity'' is understood in terms of the correlation length, because near the transition $\xi$ diverges, so there is no length scale in the system and far regions in the system are correlated. This correlation between far regions is what troubles numerical simulations to sample enough configurations and mean-field models to describe the singularity of the critical point.  

Despite the fact the phase transition can be ``visualized'' in the order parameter, the correlation length $\xi$ is the most important parameter in a phase transition, because it defines the scale of the system. As it has been pointed earlier, the scaling laws, based on the scaling hypothesis \cite{Ma} and proved by renormalization theory \cite{Wilson}, are based on the fact that the singularity of a phase transition is carried by the correlation length, and that in the critical region, the only length in the system that matters is $\xi$. Also, the finite-size scaling techniques are based on this fact. We can conclude that large spin clusters, but not the details over smaller scales, account for the physics of critical phenomena.
An example of the importance of the correlation function and the correlation length is the Berezinskii-Kosterlitz-Thouless (BKT) phase transition which we will briefly discuss in the next section.

It is indispensable to mention that given a Hamiltonian it is relatively simple to adapt the numerical algorithm in order to sample states according to the appropriate probability distribution, and that the techniques presented for recovering thermodynamic quantities are the same regardless the dimension of the system and its Hamiltonian. For example, the 3D Ising model can be solved by using the same methods presented here, where the only difference is that the number of nearest neighbors is $z=6$.

\section{Beyond the Ising model: the XY model}\label{sec::XY}

In 1966 Mermin and Wagner proved that in certain two dimensional systems,
such as the XY model, where in the Heisenberg Hamiltonian (eq. (\ref{eq:Hamiltonian})) the spin values are restricted to two dimensional unit vectors $\textbf{s}_i=(\cos \theta, \sin \theta)$, there cannot be long-range order at finite temperature \cite{Mermin}. 

What this means, is that for all $T>0$, the spin-spin correlation
function does not become constant for large separation between spins, and
the net magnetization vanishes.

However, in 1973, Kosterlitz and Thouless showed that despite the fact 
there is no long-range order, there is a phase transition at finite temperature 
in such models \cite{KT}. The phase transition is not seen neither in the 
system magnetization nor in the specific heat or the magnetic susceptibility,
but in the correlation function. For $T>T_c$ the spin-spin correlation function 
decreases exponentially, while for $T<T_c$ it falls off as a power law, 
as happens in a continuous phase transition at the critical temperature. 

The XY model exemplifies the BKT transition. It is not a continuous phase
transition as the one exhibited by the Ising model, because the order 
parameter (the magnetization) is zero for any temperature and the 
specific heat and the magnetic susceptibility are not discontinuous at the transition,
so it is referred as an essential phase transition. 

It is indeed true that there is not long-range order in the XY model, so 
Kosterlitz and Thouless proposed to call this order as topological
long-range order. The BKT transition is characterized by the existence of
vortices, which are bound in pairs of zero total vorticity (vortex buddies)
below the critical temperature, while above it they are free to move 
under the influence of a weak applied magnetic field\cite{KT}. 

\begin{figure*}[htbp]
	\centering
	\includegraphics[width=0.9\textwidth]{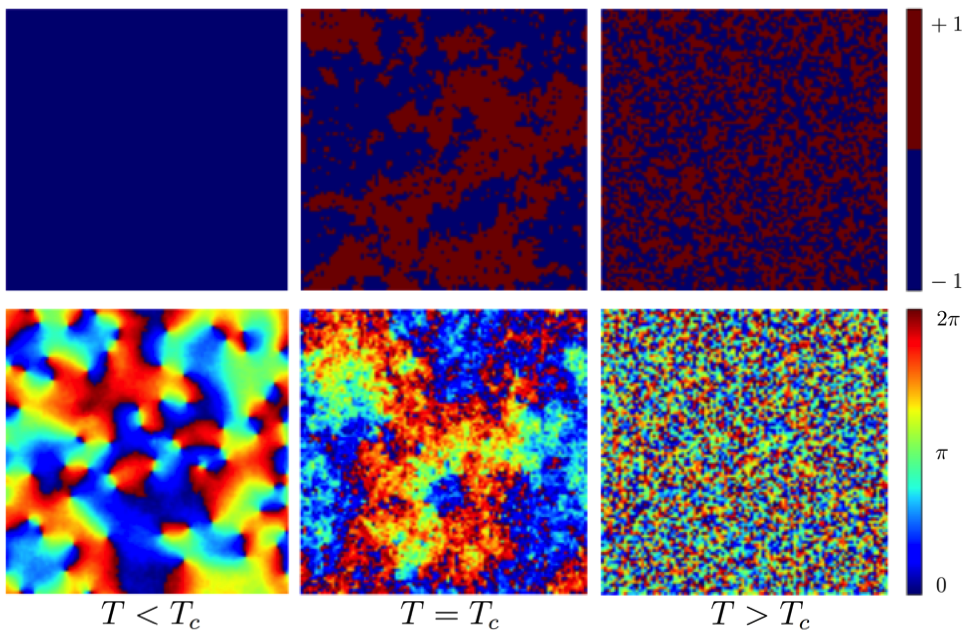}
	\caption{(Color online) Behavior of the two dimensional Ising model (top) and XY model (bottom) at $T<T_c$, $T = T_c$  and $T>T_c$. For the Ising model, an $\uparrow$ spin ($s_i=1$) is represented by red and a $\downarrow$ spin ($s_i=-1$) is represented by blue. For the XY model, because every spin is determined by $\theta$, so $\textbf{s}_i=(\cos \theta, \sin \theta)$, an angle of  $2 \pi$ is represented by red and an angle of $0$ by blue. In the low-temperature phase, the Ising model exhibits spontaneous magnetization while in the XY model, vortex buddies appear (characterized by points where a continuum from blue to red, or viceversa, circle the point. It is worth to notice that these points are present by pairs with opposite circulation). }
	\label{fig::IsingvsXY}
\end{figure*}

Vortices are the topological stable configurations in the XY model. 
Above the critical temperature unpaired vortices and anti-vortices
may be present, and below it vortex buddies formation happens
(a vortex and an anti-vortex are coupled). Thus, instead of spontaneous
magnetization of the system as happens in the Ising model, the presence
of vortex buddies characterize the BKT transition (see FIG. \ref{fig::IsingvsXY}).

The 2D XY model is interesting because it is used to describe systems in condensed
matter physics, such as superfluid helium films \cite{KT}, superconducting 
films \cite{Beasley}, Josephson junction arrays \cite{Rojas}, to cite a few examples amongst others. There is a lot of interest in explaining the phenomena previously presented and in order to do so, the BKT transition must be understood, so many efforts are focused in this problem, see for example references~\cite{Hasenbusch,Gerber_1,Gerber_2,Hsieh}.

\section{Suggested problems}\label{sec::sugg_prob}

\subsection{Implementing the Wolff algorithm}

As a first problem, the Wolff algorithm may be implemented. A nice discussion on how to do it for the Ising model is presented in references \cite{Sethna,Barkema}. The reader should compare his results with those shown in FIGS. \ref{fig::step} and \ref{fig::walk} in order to understand how different the algorithms are. The reader should be able to reproduce FIGS. \ref{fig::effects_of_size_M}, \ref{fig::effects_of_size_Chi}, \ref{fig::effects_of_size_CH} and \ref{fig::MvsH}.

If the reader is interested, he could implement the Metropolis-Hastings and the Wolff algorithm for the XY model. In the Metropolis-Hastings algorithm, each step is very similar as the one used for the Ising model, except that in each step an angle $\theta$ is proposed (instead of the flipping proposal) and $\Delta E_k$ calculated. For implementing the Wolff algorithm, see reference \cite{Wolff_DESY}.

\subsection{Higher-order spin Ising models}

The Ising model is not restricted to square lattices and spin-$1/2$ systems, but it has been extended to other geometries like triangular lattices \cite{Baxter_Yu, Ranjbar}.  Physicists have not only played with the geometry, but with the nature of the interactions and the spin angular momentum as well. 

The Blume-Emery-Grittiths (BEG) model is a spin-$1$ Ising model with a Hamiltonian given by \cite{BEG},
\begin{equation}\label{BEG}
\mathcal{H} = - J \sum_{\langle ij \rangle} \textbf{s}_i \cdot \textbf{s}_j - K  \sum_{\langle ij \rangle}  \textbf{s}_i^2 \textbf{s}_j^2 - \Delta \sum_i \textbf{s}_i^2 ,
\end{equation}
that presents a rich variety of critical and multicritical phenomena \cite{Horiguchi} and has been extended to spin-$3/2$ systems \cite{Barreto,Kaneyoshi,Bakchich,Tsushima}. The BEG model was introduced to simulate He$^3$-He$^4$ mixtures \cite{BEG}, but it has been used to describe critical phenomena magnetic systems and multi component fluids \cite{Mukamel,Lajzerowicz}.

We suggest the reader to implement the Metropolis-Hastings algorithm for the Heisenberg Hamiltonian eq.~(\ref{eq:Hamiltonian}) for a spin-$1$ or spin-$3/2$ system and calculate $E$, $M$, $C_H$, $\chi_M$ and the correlation function, then compare the results with the spin-$1/2$ system. Later, implement the BEG model and observe what does the ``multicritical phenomena'' means and how do thermodynamic properties behave in this model.

\begin{acknowledgments}
We would like to thank R. Paredes, V. Romero-Roch\'in and J. A. Seman for encouraging us to 
work in the problem, reading the manuscript and providing insightful comments, 
without their guidance and help this paper would probably not existed. E. Ibarra-Garc\'ia-Padilla would like to thank V. Romero-Roch\'in and CONACyT for the support as ``Ayudante de Investigador Nacional Nivel III'' and F.J. Poveda-Cuevas thanks SECITI-CLAF -- SECITI 064/2015 -- for a postdoctoral fellowship.
\end{acknowledgments}




\begin{thebibliography}{99}
\bibitem{footnote_extensive_var} The extensive variables of a thermodynamic system
are its energy $E$, entropy $S$, volume $V$ and particle number $N$.  
The fundamental relationships are given by the thermodynamic potentials such as 
$E=E(S,V,N)$ or the Helmholtz free energy $F=F(T,V,N)$ which are linked by 
Legendre transformations and the Euler relationship $E=TS-pV+\mu N$ that states
the extensive behavior of $E,S,V$ and $N$.
\end{thebibliography}
\end{document}